\newcommand{\hii}{H\,{\sc ii}\rm}
\newcommand{\oiii}{[O\,{\sc iii}]}
\newcommand{\oii}{[O\,{\sc ii}]}
\newcommand{\siii}{[S\,{\sc iii}]}
\newcommand{\nii}{[N\,{\sc ii}]}

\documentclass{aa}
\usepackage{graphicx}
\usepackage{txfonts}
\begin{document}
   \title{The direct oxygen abundances of metal-rich galaxies derived from electron temperature}

   \titlerunning{Oxygen abundances of metal-rich galaxies from $T_e$}

   \author{Y. C. Liang\inst{1,2}, F. Hammer\inst{2}, S. Y. Yin\inst{1,3,4}, H.
   Flores\inst{2}, M. Rodrigues\inst{2}, Yanbin Yang\inst{2,1}}

    \authorrunning{Liang et al.}

   \offprints{Y. C. Liang: ycliang@bao.ac.cn}

   \institute{
  $^1$National Astronomical Observatories, Chinese Academy of Sciences,
     20A Datun Road, Chaoyang District, Beijing 100012, China\\
  $^2$GEPI, Observatoire de Paris-Meudon, 92195 Meudon, France \\
  $^3$Department of Physics, Hebei Normal University,Shijiazhuang 050016, China \\
  $^4$Department of Physics, Harbin University, Haerbin 150086, China
 }

 \date{Received; accepted 17 July 2007}


  \abstract
   {Direct measurement of oxygen abundance for metal-rich galaxies from
   electron temperature is difficult or impossible since 
   temperature-sensitive auroral lines generally become too weak to be
   measured.} 
   {We aim to derive the electron temperature ($T_e$)
   in the gas of metal-rich star-forming galaxies, which can be obtained from
   their ratios of auroral lines [O~{\sc ii}]$\lambda\lambda$7320,7330
   to nebular lines [O~{\sc ii}]$\lambda$3727, in order to establish a more
   robust mass-metallicity relationship, and compare the $T_e$-based (O/H)
   abundances with those from empirical strong-line calibrations, 
   such as $R_{23}$ (=([O~{\sc ii}]$\lambda$3727+[O~{\sc
   iii}]$\lambda\lambda$4959,5007)/H$\beta$).
  }
   {We obtained 27 spectra by stacking the spectra of several hundred
   (even several thousand) star-forming galaxies selected from the SDSS-DR4
   in each of the 27 stellar mass bins from log($M_*$)$\sim$ 8.0 to 10.6
   (in units of log($M_{\odot})$).
   This ``stack" method sufficiently improves the signal-to-noise ratio
    of the auroral lines [O~{\sc ii}]$\lambda\lambda$7320,7330. Using a
   two-zone model for the temperature structure, we derive the electron
   temperature $t_2$ in the low ionization region
   from the [O~{\sc ii}]$\lambda\lambda$7320,7330/[O~{\sc ii}]$\lambda$3727 ratio,
   and then use a relation derived by fitting H~{\sc ii} region models
   to estimate the electron temperature $t_3$ in the high ionization region
   from $t_2$. Then, the  direct (O/H) abundances
   are obtained from $t_2$, $t_3$ and the related line-ratios.
   The emission lines have been carefully corrected for dust extinction 
   using the Balmer line ratio after correcting for the underlying
   stellar absorption.
   }
   {Combining our results with those from the literature with lower metallicities,
   we are able to provide a new relationship between  the direct
measurements of (O/H) and $R_{23}$, which still shows an upper and
a lower branch with the transition around
12+log(O/H)~$\sim$8.0-8.2. It also shows that the empirical $R_{23}$ method will
overestimate log(O/H) by 0.2 to 0.6 dex. The new metal-mass
relationship of the galaxies with moderate metallicities 
is fitted by a linear fit (12+log(O/H) =
6.223+0.231$\times$log($M_*$)) confirming that empirical methods
significantly overestimate (O/H). We also derived their (N/O) abundance
ratios on the basis of the $T_e$ method, which are consistent with
the combination of the primary and secondary components of
nitrogen.
    }
   {This study provides for the first time a method to calibrate direct O/H
abundances (from $T_e$) for a large range of galaxies within a stellar mass
range of $\sim$5\,10$^8$$M_{\odot}$  to 4\,$10^{10}$$M_{\odot}$.}
   \maketitle
%

%
\section{Introduction}
 The chemical properties of stars and gas within a galaxy
provide both a fossil record of its star formation history
and information on its present evolutionary status.
 Accurate abundance
measurements for the ionized gas in galaxies require the
determination of electron temperature ($T_e$) in the gas, which is
usually obtained from the ratio of auroral to nebular line
intensities, such as [O~{\sc iii}]$\lambda\lambda4959,5007$/[O~{\sc
  iii}]$\lambda$4363.  This is generally known as the
``direct $T_e$-method" since $T_e$  is
  directly inferred from observed line ratios. However, it is well known that
  this procedure is difficult to carry out for metal-rich galaxies
  since, as the metallicity increases, the electron temperature
decreases (as the cooling is via metal lines), and the auroral lines
eventually become too faint to measure.
Instead, other strong nebular line ratios have to be used to
estimate the oxygen abundances of metal-rich galaxies
(12+log(O/H)$\geq$8.5), such as the $R_{23}$ (=([O~{\sc
  ii}]$\lambda$3727+[O~{\sc iii}]$\lambda\lambda$4959,5007)/H$\beta$)
parameter (Pagel et
al. 1979; Tremonti et al. 2004, hereafter T04, and the references therein),
 N2 index (=log([N~\textsc{ii}]$\lambda $6583/H$\alpha $)),
and some other
metallicity-sensitive strong-line ratios
(Kewley \& Dopita 2002; Pettini \& Pagel 2004;
Stasinska 2006; Liang et al. 2006; Yin et al. 2007a).
Pilyugin (2001a,b) and Pilyugin \& Thuan (2005)
 also suggested their $P$-method to estimate the (O/H) abundances from the
 $R_{23}$ and $P$ (=[O~{\sc iii}]/([O~{\sc ii}]+[O~{\sc iii}])) parameters.

However, these ``indirect" calibrations may produce some errors
in metallicity estimates.
For example,  the
double-valued problem of $R_{23}$ for oxygen abundances, with the transition
occurring near 12+log(O/H)$\sim $8.4 (log$R_{23}$$ \sim $0.8) 
(McGaugh 1991).
Most of the other strong-line ratios are sensitive to the
photoionization parameter in the ionized gas (Kewley \& Dopita 2002),
and some of them suffer from the same double-valued problem
as $R_{23}$.

Researchers continuously try to obtain the
``direct" $T_e$-based oxygen abundances for the metal-rich
galactic gas, though there are some disagreements (Stasinska 2005). For
example, Kennicutt et al. (2003; hereafter KBG03), Bresolin et al.
(2004, 2005), Bresolin (2007) and Garnett et
 al. (2004a,b) obtained the oxygen abundances from $T_e$ for
 a handful of H~{\sc ii} regions in M101 and M51
 with 12+log(O/H) $>$ 8.4 by using  high quality spectra taken from the
 Very Large Telescope (VLT),  
 Multiple Mirror Telescope (MMT) or Keck, 4-8m telescopes.
 Pilyugin et al. (2006) derived a flux-flux(ff)-relation to estimate
 the [O~{\sc iii}]$\lambda$4363/[O~{\sc iii}]$\lambda$5007 ratio from the
 obtained $R_{23}$ by using a sample of 48
 H~{\sc ii} regions with moderate metallicities.
However, all these impressive progresses are only based on
 individual H~{\sc ii} regions.
 It is very difficult to measure the temperature-sensitive auroral lines from
 the integrated spectra of metal-rich galaxies,
 hence derive their ``direct" metallicities.

 The wonderful Sloan Digital Sky Survey (SDSS) database 
 allows this to be realized.  It
 provides high quality spectra
 covering 3800-9200\AA~
 with moderate resolution ($\sim$3\AA)
 for a large number of galaxies at different given oxygen
 abundances or stellar masses.
 It makes it possible for us to
 measure the temperature-sensitive auroral lines
 [O~{\sc ii}]$\lambda\lambda$7320,7330 from the integrated light of 
 metal-rich galaxies, which are usually the
strongest auroral lines in the measured optical spectra of the metal-rich galactic gas.
 To do so, we combine the spectra of
 several hundred (even several thousand) galaxies in each of the different mass bins (see
 Sect.~\ref{sec2} for details).
  This ``stack" method significantly improve the signal-to-noise (S/N) ratio
   of the spectra, which makes it possible to use
    [O~{\sc ii}]$\lambda\lambda$7320,7330/[O~{\sc ii}]$\lambda$3727 ratios
   to derive their electron temperature $T_e$ in the low ionization
   region ($t_2$), then to obtain their ``direct" (O/H) abundances.
   This is probably the first result concerning the derivation 
   of ``direct" (O/H) abundances from $T_e$
   from the integrated light of metal-rich star-forming galaxies.

    This paper is organized as follows. The sample selection criteria
    and spectral combination are
described in Sect.\ 2. The corrections for stellar absorption and flux measurements
are described in Sect.\ 3.
In Sect.\ 4, we re-calibrate the relations of the (O/H) abundances
versus stellar masses and $R_{23}$ parameters
for star-forming galaxies on the basis of the SDSS-DR4 database.
The determinations of the oxygen abundances
from $T_e$ are presented in Sect.\ 5.  In Sect.\ 6, we
compare the $T_e$-based (O/H) abundances with those
derived from other empirical strong-line methods, such as
$R_{23}$, $P$ and N2, and derive a new relationship between (O/H)$_{T_e}$
and $R_{23}$. In Sect.7, we obtain
the log(N/O) abundance ratios of these combined galaxies.
The discussions are given in Sect.\ 8. We conclude the paper in Sect.\ 9.
Throughout this paper, the stellar masses of galaxies $M_*$
are in units of solar mass $M_{\odot}$.
[O~{\sc ii}]$\lambda\lambda$7320,7330 can also be given as one single line
[O~{\sc ii}]$\lambda$7325, so both of these two descriptions 
will appear in the text.

\section{Sample selection and spectral combination}
\label{sec2}

We aim to measure the electron temperatures,
hence the $T_e$-based (O/H) abundances,
for massive galaxies. 
[O~{\sc ii}]$\lambda\lambda$7320,7330 lines are usually the
strongest auroral lines in the measured optical spectra of
star-forming galaxies (Bresolin et al. 2005),
thus we focus on the $T_e$ derived from the line-ratio
of [O~{\sc ii}]$\lambda\lambda$7320,7330/[O~{\sc ii}]$\lambda$3727.
To increase the S/N ratio of the spectra, we combine
the spectra of several hundred (even several thousand) galaxies
selected following the criteria described below,
which makes it possible to measure
[O~{\sc ii}]$\lambda\lambda$7320,7330 auroral lines for the massive galaxies
 up to log($M_*$)$=$10.6.

Firstly, we select $\sim$70,000 star-forming galaxies with
8.0$<$log($M_*$)$<$10.6 from SDSS-DR4
by following their
stellar masses and metallicities estimated by
the MPA/JHU group\footnote{http://www.mpa-garching.mpg.de/SDSS/}
(T04; Kauffmann et al. 2003; Brinchmann et al. 2004;
also see Liang et al. (2006) for the criteria for star-forming galaxies).
 This criterion allows us to select all 
 the star-forming galaxies with available estimates of stellar masses and metallicities 
 (above the required S/N ratios) from 
 the SDSS-DR4, but with stellar masses smaller than 4 $10^{10}M_{\odot}$.
 This mass limit is caused by the difficulty in obtaining measurements of 
 [O~{\sc ii}]$\lambda\lambda$7320,7330 auroral lines for the most massive galaxies. 

Secondly, we divide these galaxies into 27 bins with different stellar masses
from log($M_*$)=8 to 10.6.
The mass bins and the total numbers of the sample galaxies in the bins
are given in Col.~3 and Col.~5 in Table~\ref{tab1}, respectively.

Thirdly, in each bin, we select about half of the galaxies
with stronger emission lines
as working samples, which allows us to obtain
[O~{\sc ii}]$\lambda\lambda$7320,7330 lines at a high S/N ratio for the galaxies
with stellar masses up to log($M_*$)$=$10.6.
Namely, we select the half of the galaxies having
equivalent width values of
[O~{\sc ii}]$\lambda$3727, W(O~{\sc ii}), larger than the mean value of the
whole sample of galaxies in each bin for the case of log($M_*$)$<$10.0,
and larger than 30\AA~ for the case of log($M_*$)$\geq$10.0.
The  W(O~{\sc ii}) of the selected
galaxies and the mean W(O~{\sc ii})$_{\rm mean}$ (Col.~4 in Table~\ref{tab1},
given as the actual EW([O~{\sc ii}]))
in each bin were gathered or
calculated from the measurements published
by the MPA/JHU group.
The numbers of the selected samples in the bins
are given in Col.~6 in Table~\ref{tab1}.
The selected sample for this study finally consists of
23,608 galaxies. 
In the following (see Sect.~\ref{sec4}), we will show that the above criterion
creates almost no bias in estimating the mass-metallicity relationship.

Fourthly, the hundred (even several thousand) spectra of the sample galaxies
in each bin are combined to be one single spectrum, which
has high S/N ratio ($\sim$60 at continuum) and sufficiently strong 
[O~{\sc ii}]$\lambda\lambda$7320,7330 lines for measuring.
Figure~1 shows all the combined spectra
at rest-frame in 27 mass bins, and the small figure in each
panel shows the auroral lines [O~{\sc ii}]$\lambda$$\lambda$7320,7330
region.
The basic data of the combined spectra are given in Table~\ref{tab1}.

These combined spectra may provide a good representation of the integrated
light of the real galaxies in a given stellar mass range. Obviously, they
gather more light from more individual H~{\sc ii} regions of the galaxies with
similar properties, e.g., the similar stellar masses. However, one may wonder
whether stacked spectra may keep the information that underlines the real
physics of each of the individual galaxies.  
Indeed the intrinsic property of this is similar to what happens
when studying the global properties of an individual galaxy: it is a
mix, mainly of the H~{\sc ii} regions of the galaxy. Here we simply sum up a
much larger number of H~{\sc ii} regions in a large number of galaxies, and
assume that the resulting properties are near the average properties of each
individual galaxy. In  Sect.~\ref{sec4} we will verify that this assumption
indeed applies for the empirical determination of O/H (through the so-called
$R_{23}$ method), and the stacking method  will be discussed further in
Sect.\ref{sec8}.2.

From our integrated-light spectra, it is still almost impossible to measure
other temperature-sensitive emission-lines [N~{\sc ii}]$\lambda$5755,
[S~{\sc ii}]$\lambda$4072.
And the SDSS spectrum does not
cover [S~{\sc iii}]$\lambda$$\lambda$9069,9532
though it is possible to measure the
[S~{\sc iii}]$\lambda$6312 emission-line for the sample with
log($M_*$)$<$9.6. Thus it is impossible
to obtain directly the  $T$(\siii) from the [S~{\sc iii}] ratio
and the $T$(\nii) from the [N~{\sc ii}] ratio
for these sample galaxies. In a two-zone model of temperature structure,
$T(N^+)$ is equal to $T(O^+)$ and it is acceptable to omit
  $T(S^{+2})$ since it represents an intermediate-ionization zone.

\begin{figure*} [b]
\centering
\input epsf
\epsfverbosetrue
\epsfxsize 13.8cm
\epsfbox{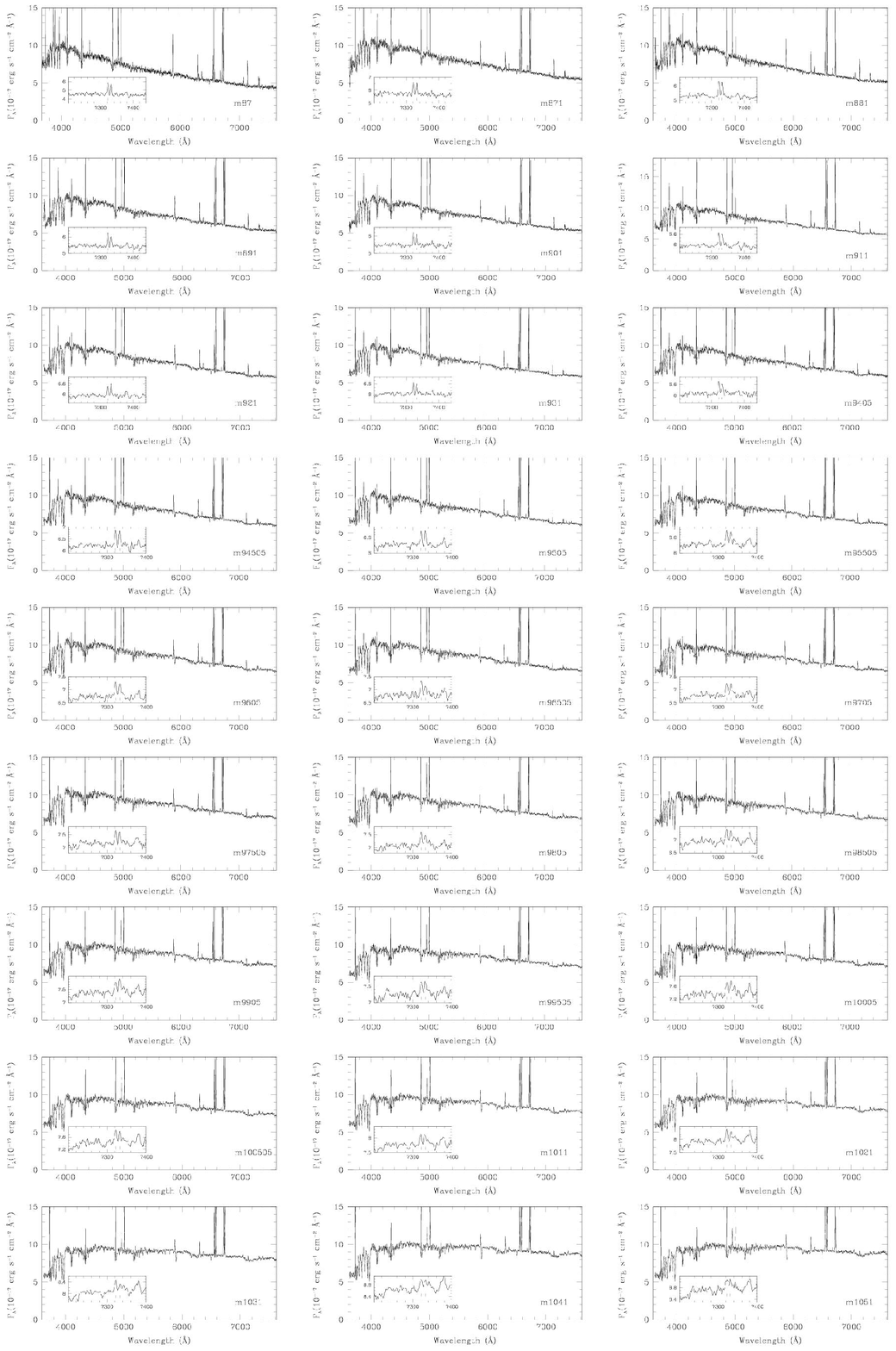} 
\caption {The 27 combined spectra of the
sample galaxies at rest-frame in the corresponding mass bins. 
The small figure inside each panel
shows the auroral lines [O~{\sc ii}]$\lambda$$\lambda$7320,7330.
 The label for the mass bin of each spectrum is given
in the bottom right corner.
}
\label{fig1}
\end{figure*}

\begin{table}
\begin{center}
\caption{\footnotesize
The basic properties of the 27 combined spectra.
Col.~(1)-(6) refer to the counting number, 
the label for the mass bin (marked in Fig.~1), 
the mass bin, the average
EW([O~{\sc ii}]$\lambda$3727) of the sample galaxies in each
bin, the total number of the galaxies, and
the number of the selected spectra to be combined (about half) in each bin,
respectively.
 }
\label{tab1}
\scriptsize
\begin{tabular}{llllrr}
\hline
\hline
\noalign{\smallskip}
(1)  & (2) & (3) & (4) & (5) & (6) \\ \hline
\noalign{\smallskip}
No. & Label  & Mass bin      & EW([O~{\sc ii}]) & Total  & Selected     \\
    &      & (log$M_*$)     & average (\AA)       & Number & Number   \\ \hline
1  & m87   & $<$ 8.7          &  -78.72     &     378        &  149  \\
2  & m871  & 8.7-8.8          &  -73.73     &     339        &  158 \\
3  & m881  & 8.8-8.9          &  -71.53     &     533        &  228  \\
4  & m891  & 8.9-9.0          &  -65.51     &     797        &  341  \\
5  & m901  & 9.0-9.1          &  -62.24     &     1227       &  533   \\
6  & m911  & 9.1-9.2          &  -60.42     &     1470       &  593 \\
7  & m921  & 9.2-9.3          &  -57.76     &     1885       &  766  \\
8  & m931  & 9.3-9.4          &  -55.02     &     2428       & 1029  \\
9  & m9405 & 9.40-9.45        &  -52.80     &     1329       &  552 \\
10 & m94505 & 9.45-9.50       &  -51.31     &     1478       &  627  \\
11 & m9505 & 9.50-9.55        &  -50.41     &     1670       &  705  \\
12 & m95505 & 9.55-9.60       &  -48.14     &     1746       &  751 \\
13 & m9605 & 9.60-9.65        &  -46.58     &     1920       &  814  \\
14 & m96505 & 9.65-9.70       &  -45.11     &     2001       &  837  \\
15 & m9705  & 9.70-9.75       &  -43.72     &     2156       &  906 \\
16 & m97505 & 9.75-9.80       &  -42.17     &     2262       &  951 \\
17 & m9805  & 9.80-9.85       &  -40.21     &     2428       & 1010  \\
18 & m98505 & 9.85-9.90       &  -38.31     &     2841       & 1020 \\
19 & m9905  & 9.90-9.95       &  -35.94     &     2552       & 1025  \\
20 & m99505 & 9.95-10.0       &  -33.59     &     2605       & 1048  \\
21 & m10005 & 10.0-10.05      &  -33.22     &     2750       & 1076  \\
22 & m100505 & 10.05-10.10    &  -31.19     &     2814       & 1120 \\
23 & m1011   & 10.10-10.20    &  -29.12$^a$  &     5860       & 2145  \\
24 & m1021   & 10.20-10.30    &  -26.28$^a$  &     6004       & 1813 \\
25 & m1031   & 10.30-10.40    &  -23.79$^a$  &     6084       & 1417 \\
26 & m1041   & 10.40-10.50    &  -22.18$^a$  &     5973       & 1148 \\
27 & m1051   & 10.50-10.60    &  -20.72$^a$  &     5341       &  846  \\  \hline
\end{tabular} 
\end{center}
$Note ~(a): $ for the massive galaxies with log($M_*$)$>$10.10,
we select those with
stronger [O~{\sc ii}] emission-lines with EW([O~{\sc ii}])$<$ -30\AA~ 
for spectral combination.
\end{table}

\section{Correction for stellar absorption and extinction}

 The Balmer emission lines
should be corrected carefully by the underlying stellar absorption
 before measuring their fluxes properly. This correction
dominates the main uncertainties for the flux measurements
of the Balmer lines in this study since the S/N ratios
of the combined spectra are quite high.

\subsection{Stellar absorption correction}
\label{sec3.1}

 The methodology used here is fitting an observed spectrum, including its continuum
 and absorption lines, with a linear
 combination  of simple stellar populations (SSP) generated using the
 population synthesis code of Bruzual \&
Charlot (2003, BC03), GALAXEV. The BC03 models incorporate an empirical spectral library
 with a wavelength coverage
(3200-9500\AA) and spectral resolution ($\sim$3\AA) that is well matched to that
of the SDSS data. We use 10 templates of SSP of BC03 with different ages
(0.005, 0.025, 0.1, 0.2, 0.6, 0.9, 1.4, 2.5, 5, 10 Gyr) at solar metallicity
that are obtained by using the stellar evolution tracks of the ``Padova
1994 library", and the Chabrier's IMF (Chabrier 2003).
These 10 templates are a sub-sample of the 39 templates used by
T04 to fit the spectra of the SDSS galaxies, which
consist of three metallicities $Z=1/5, 1, 2.5Z_{\odot}$.
The 10 templates with $Z_{\odot}$ used in this study
should be sufficient to correct the
stellar absorption for the spectra, since the degeneracy of age and metallicity still
cannot be perfectly broken.

\begin{figure}
\centering
\input epsf
\epsfverbosetrue \epsfxsize 8cm
\epsfbox{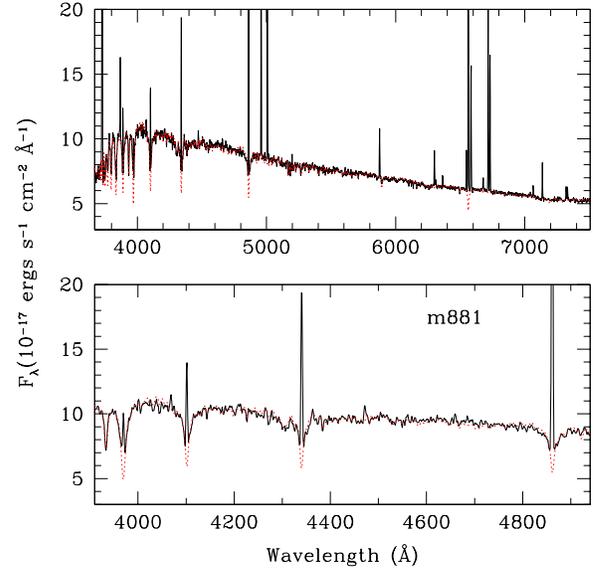}
\caption{The fitting on the continuum and stellar absorption for
one example spectrum m881. The top panel is for the spectral range
from 3670 to 7505 \AA, and the bottom panel is for the detailed
fitting from Ca H K lines to H$\beta$ 
(from 3910 to 4940 \AA). The solid lines indicate
the observed spectrum, and the dotted lines give the fittings by
using the SSP of BC03 and the program STARLIGHT (see Sect.~\ref{sec3.1}).
The top and bottom plots are made by using the same data and model.   }
\label{fig2}
\end{figure}

We use the program, STARLIGHT\footnote{http://www.starlight.ufsc.br}, 
developed by Cid Fernandes and colleagues,
to synthesize the stellar absorptions and continua of the sample galaxies 
(Cid Fernandes et al. 2005, 2007; Mateus et al. 2006,  Asari et al. 2007).
It is a program to fit an observed spectrum $O_\lambda$ with a model $M_\lambda$
that adds up $N_*$ (we use 10 as mentioned above) spectral
components from a pre-defined set of base spectra. 
Both dust extinction on the SSP templates and line-of-sight stellar motions
modeled by a Gaussian distribution with dispersion $\sigma$ have been considered
in the program.
The Galactic extinction law of Cardelli et al. (1989, CCM) with $R_V$=3.1 is adopted. 
The fit is carried out with the Metropolis scheme,
 which searches for the minimum $\chi^{2}$.
 We show an example in Fig.~\ref{fig2} for the spectral fittings.

\subsection{Flux measurements and extinction correction}
\label{sec3.2}

\begin{figure}
\centering
\input epsf
\epsfverbosetrue \epsfxsize 7.8cm
\epsfbox{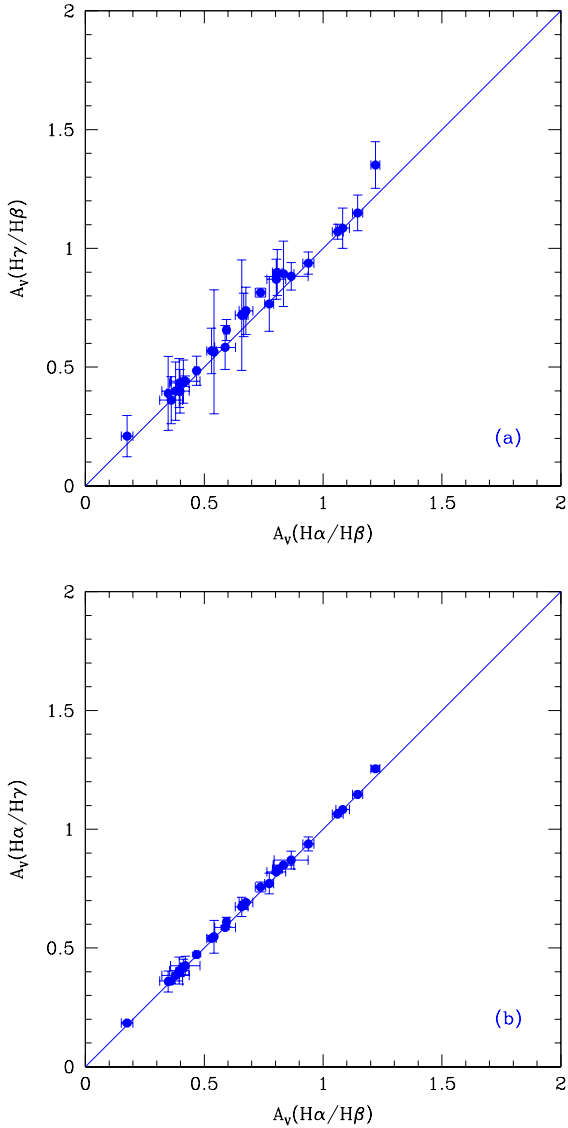}
\caption{Comparisons between the extinction values derived from
H$\alpha $/H$\beta $, H$\gamma $/H$\beta $ and H$\alpha $/H$\gamma $ for the sample
galaxies. $A_V$(H$\alpha $/H$\beta $) is adopted as the actual
extinction in this study. 
              }
\label{fig3}
\end{figure}

We use SPLOT in the IRAF\footnote{IRAF is distributed by the National Optical Astronomical
Observatories, which is operated by the Association of Universities for Research
in Astronomy, Inc., under cooperative agreement with the National Science
Foundation.} package to measure the fluxes of the emission-lines from
the stellar-absorption subtracted spectra.
Dust extinction ($A_V$, at V-band),
is estimated by using the Balmer-line ratios
H$\alpha $/H$\beta $, H$\gamma $/H$\beta $ and H$\alpha $/H$\gamma $,
and by assuming case B
recombination with a density of 100\thinspace cm$^{-3}$, a temperature
of 10$^{4}$\thinspace K, and the predicted intrinsic H$\alpha $/H$\beta $ ratio of 2.86 
and H$\gamma $/H$\beta $ ratio of 0.466 (Osterbrock 1989).
Since the stellar absorptions underlying Balmer emission-lines
have been corrected carefully by using the
synthesized SSP,
the dust extinction values derived from the three line ratios
should be almost identical.
Figure~\ref{fig3} shows the excellent consistency of the derived values for the dust
extinction. 

The uncertainties of the flux measurements are dominated by the
underlying stellar absorption correction.  Another source comes
from the independent measurements by two us us, YCL and SYY. 
Figure~\ref{fig3} shows that the error-bars of $A_V$(H$\gamma $/H$\beta $)
are moderately larger than the other two from H$\alpha $/H$\beta $ and H$\alpha $/H$\gamma $, simply
because the H$\gamma $ line is weaker than the other two lines.
H$\alpha $ is the strongest of these three lines, and
is affected less than the other two by the stellar absorption.
In the following, we adopt $A_V(H\alpha/H\beta)$ as
the actual extinction inside the galaxies.

\section{Consistency checks of the empirical O/H estimates using the $R_{23}$ method }
\label{sec4}

It is necessary to re-derive the relationship between (O/H) abundances
 and stellar masses and $R_{23}$ parameters
for nearby star-forming galaxies on the basis of the SDSS-DR4 database
by following the method of T04
since our sample galaxies are selected from DR4.
This revises somewhat the calibrations obtained from the DR2 database by T04,
but should be more reasonable for this study.

\begin{figure}
\centering
\input epsf
\epsfverbosetrue \epsfxsize 6.6cm
\epsfbox{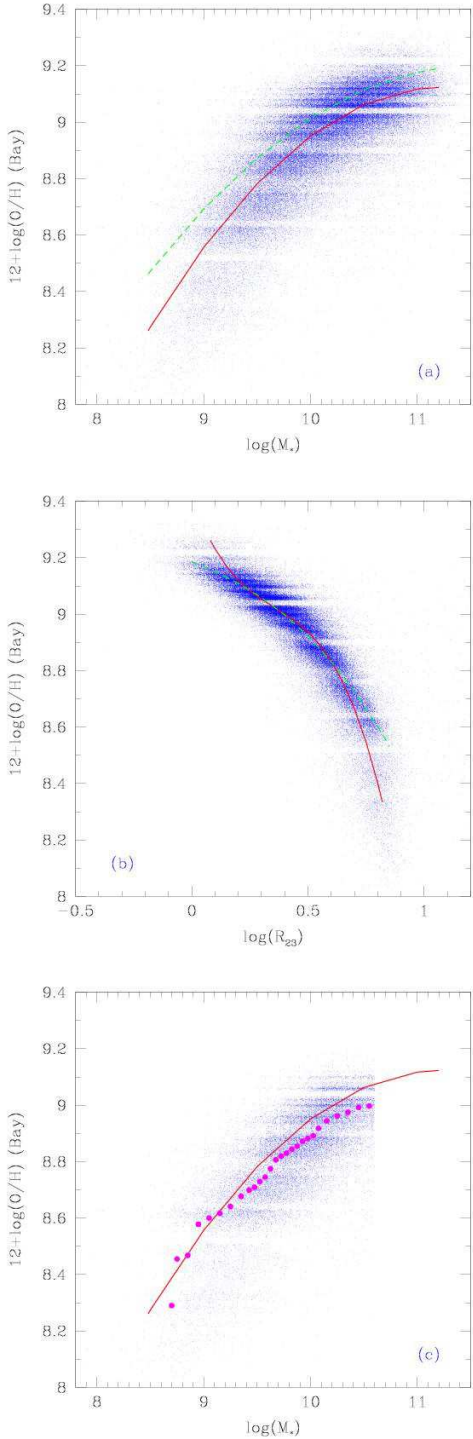}
\caption{Relation of the log(O/H) abundances versus the stellar
masses (a) and versus the log($R_{23}$) (b) of the nearby
star-forming galaxies selected from SDSS-DR4 (the small points,
$\sim$85,000). The two solid lines in (a) and (b) refer to the new calibrations
derived from the SDSS-DR4 star-forming galaxies with
8.5$\leq$log($M_*$)$\leq$11.3 (Eq.~(\ref{eqOHmass})), 
and 8.3$\leq$12+log(O/H)$\leq$9.2 (Eq.~(\ref{eqOHR23})),
respectively. The dashed lines refer to the calibrations obtained
by T04 from DR2 database.
(c) represents
the sample of galaxies selected for this study.  
The large filled circles refer to the 
median-value points in the mass bins following Table~\ref{tab1},
which are very close to the solid line,
the same one as in Fig.~\ref{figdr4}a. 
(Please see the online color version of the plots for the lines.)
      }
\label{figdr4}
\end{figure}

Figure~\ref{figdr4}a,b shows the relationship between 
12+log(O/H) and log($M_*$) and log($R_{23}$)
for the DR4 star-forming galaxies
($\sim$85,000, the small points). The dashed lines are the calibrations of
T04 obtained from DR2, and the solid lines are the re-derived relations from DR4, given as
the following formulas:
\begin{equation}
\rm 12+log(O/H) = -5.292 + 2.568(logM_*) -0.11441(logM_*)^2,
\label{eqOHmass}
\end{equation}
\begin{eqnarray}
\rm 12+log(O/H) & = & 9.412 - 2.251(\rm logR_{23}) + 4.885(\rm logR_{23})^2  \nonumber \\
                & - & 4.564(\rm logR_{23})^3.
\label{eqOHR23}
\end{eqnarray}
They are obtained by fitting the median values
in the bins of
0.1\,dex in log($M_*$) and 0.05\,dex in log(O/H), respectively.

Figure~\ref{figdr4}a,b reveals the differences between the results from DR4 and DR2
for these two relationships which might be due to the expansion of the database.
Eq.~(\ref{eqOHmass}) will result in a lower log(O/H) abundance
than the previous calibration
at the given stellar mass, generally about 0.1\,dex.
Eq.~(\ref{eqOHR23}) will result in a higher and lower log(O/H) abundance
than the previous one
at the given log($R_{23}$) at the metal-rich and metal-poor ends, respectively.

Figure~\ref{figdr4}c shows
our selected galaxies (the small blue points) that
we extracted from the whole sample
of SDSS-DR4 galaxies following the criteria described in Sect.~\ref{sec2}.
The large solid circles refer to the 
median-value points in the 27 mass bins given in Table~\ref{tab1}.
The solid line is the same as in Fig.~\ref{figdr4}a and 
Eq.~(\ref{eqOHmass}).
It shows that these median-value points of the selected samples 
are very close to the median-value relation of the whole sample (the solid line).
Notice however that it slightly biases O/H values towards lower metallicities 
at the highest mass end.

\begin{figure}
\centering
\input epsf
\epsfverbosetrue \epsfxsize 7.8cm
\epsfbox{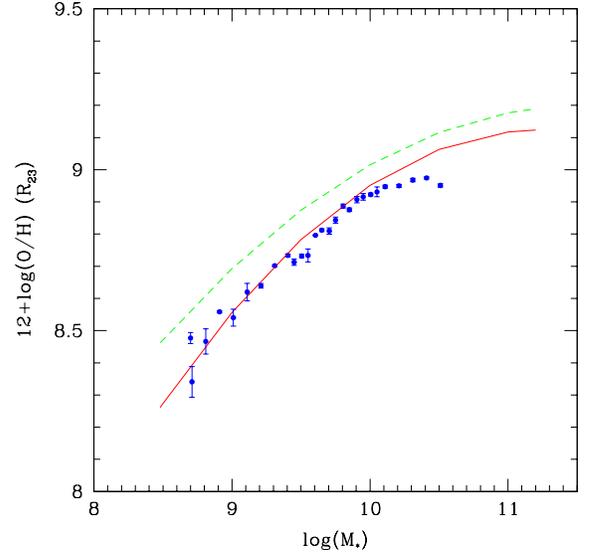}
\caption{Relationship between the
12+log(O/H)$_{R_{23}}$ abundances and the stellar masses for our
27 combined galaxies (the points). The solid and dashed lines are the
same as in Fig.~\ref{figdr4}a.
      }
\label{fig4}
\end{figure}

Figure~\ref{fig4} presents the 12+log(O/H)$_{R_{23}}$ versus log($M_*$)
relationship of our 27 stacked spectra.
Their 12+log(O/H)$_{R_{23}}$ abundances are estimated
from the extinction-corrected $R_{23}$ parameters
following Eq.~(\ref{eqOHR23}).
The corresponding stellar mass in each bin is the middle value
there (see Table~\ref{tab1}).
Notice that the data points obtained from
the stacked spectra are very similar to the large filled circles 
in Fig.~\ref{figdr4}c,
i.e., the actual median-value points of the sample galaxies in the same mass
bins. Moreover, the stacked spectra provide values quite close to the solid line,
which represents the median-value relation of the whole sample. This implies
that for the empirical O/H estimate using the $R_{23}$ method, our stacking
method provides a good representation of the median value for the whole sample,
and that the stacking method is able to give a reasonable estimate of the
behavior of the global mass-metallicity relationship.

\section{The ``direct" oxygen abundances derived from electron temperature $T_e$}

A two-zone model for the temperature structure within the galactic gas
was adopted. In this model, $T_e$([O~{\sc ii}]) (in K, as $t_2$ in 10$^{4}$K)
represents the temperature for
low-ionization species such as $O^{+}$ (and $N^{+}$),
while $T_e$([O~{\sc iii}]) (in K, as $t_3$ in 10$^{4}$K)
represents the temperature for high-ionization species such as
$O^{++}$.

We estimate $t_2$ from the ratio
of [O~{\sc ii}]$\lambda\lambda$7320,7330 to
[O~{\sc ii}]$\lambda$3727 by using the task {\sl temden.nebular} in the
{\sc IRAF/STSDAS} package
(de Robertis et al. 1987; Shaw \& Dufour 1995);
 $t_3$ in the
high-ionization zone is determined from $t_2$ following an
equation derived by fitting H {\sc ii} region models. Several versions
of the $t_2$-$t_3$ relationship have been proposed (Pilyugin et al. 2006).
We use the one given by Garnett (1992), which is valid over the range
2000K$<$$T_e$([O~{\sc iii}])$<$18,000K, and has been widely used:
\begin{equation}
t_2 = 0.7t_3 + 0.3.
\end{equation}

We can then obtain the ``direct" oxygen abundances from
electron temperature
for these combined galaxies by using the equations published
by Izotov et al. (2006) for
the determination of the oxygen abundances in H {\sc ii} regions for a
five-level atom. They used the atomic data from the references listed
in Stasi\'{n}ska (2005). The formulas are
\begin{eqnarray}
12 + {\rm log}(O^{+}/H^+) =
{\rm log}(I_{{[O}{II]\lambda3726+\lambda3729}}/I_{H\beta}) +  \nonumber \\
      5.961 + \frac{1.676}{t_2} - 0.40{\rm log} t_2 - 0.034 t_2 +  \nonumber \\
      {\rm log}(1 + 1.35x_2),
\label{eqo2}
\end{eqnarray}
where $x_2 = 10^{-4}n_et_2^{-1/2}$,
and $n_e$ is the electron density in cm$^{-3}$, and
\begin{eqnarray}
12 + {\rm log}(O^{++}/H^+) =
{\rm log}(I_{{[O}{III]\lambda4959+\lambda5007}}/I_{H\beta}) + \nonumber \\
                 6.200 + \frac{1.251}{t_3} - 0.55{\rm log}t_3 - 0.014t_3.
\end{eqnarray}
The total oxygen abundances are derived from the following
equation:
\begin{equation}
\frac{O}{H} = \frac{O^+}{H^+} + \frac{O^{++}}{H^+}.
\end{equation}

The electron densities in the ionized gas of the galaxies are
calculated from the line ratios [S~{\sc ii}]$\lambda6717$/[S~{\sc
  ii}]$\lambda6731$ (Cols.~2,3 in Table~2) by using the
  five-level statistical equilibrium
model in the task {\sl temden.nebular} in the {\sc IRAF/STSDAS}
package at $T_e$=10,000K.  Notice that
$x_2$ in Eq.~(\ref{eqo2}) has a very small impact since
it is generally less than 0.1 with $n_e$ $<$ 10$^3$ cm$^{-3}$.

Table~2 lists the derived properties of the combined galaxies,
including the mass-bin, the [S~{\sc ii}]$\lambda6717$/[S~{\sc
  ii}]$\lambda6731$  ratios, the electron density,
  the [O~{\sc ii}]$\lambda\lambda7320,7330$/[O~{\sc ii}]$\lambda3727$  ratios,
  the electron temperature $T_e$([O~{\sc ii}]) (in K), $T_e$([O~{\sc iii}]) (in
  K), and the $T_e$-based 12+log(O/H) abundances, 
  which are approximately 8.0 to 9.0.

\begin{figure}
\centering
\input epsf
\epsfverbosetrue \epsfxsize 7.8cm
\epsfbox{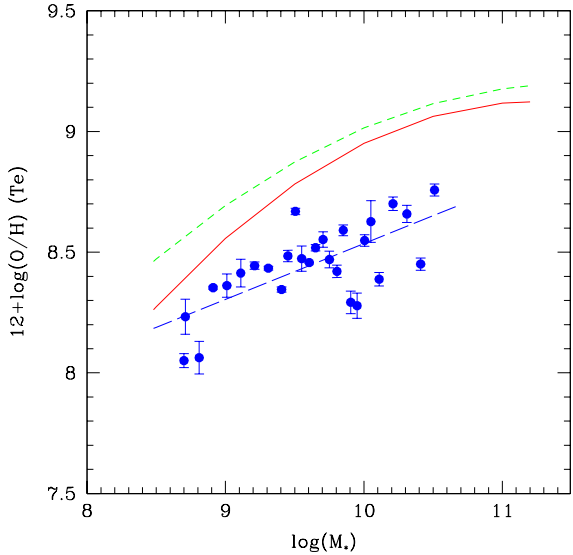}
\caption{Relationship between the (O/H) abundances derived from $T_e$
 and the stellar masses of the combined galaxies.
 The long-dashed line is the linear least-square fit
  for the data points.
  The solid and short-dashed lines are the same as in Fig.~\ref{fig4}.
           }
\label{figTeR23}
\end{figure}

 Figure~\ref{figTeR23} shows the relationship between the
 $T_e$-based log(O/H) abundances and the stellar masses log($M_*$)
 for the combined galaxies (the filled circles).
 The thick long-dashed line is the
  linear least-square fit
  for the data point and is given by
 \begin{equation}
 \rm 12+log(O/H) = 6.223 + 0.231 \times (logM_*),
 \label{eq5}
 \end{equation}
 with an rms about 0.12\,dex, which shows the increasing log(O/H)
 following the stellar masses. 

  This shows that the $T_e$-based log(O/H) abundances of the upper branch are
  generally about 0.2-0.4\,dex (down to 0.6\,dex) lower
  than the $R_{23}$-based abundances (the solid and dashed lines,
  the same as in Fig.~\ref{fig4}) at a given stellar mass. The error-bars
  marked here come from the flux measurements. We did not consider
  the uncertainties from the conversion relationship between $t_2$ and $t_3$,
  and the temperature fluctuations and gradients (see the discussions in
  Sect.~\ref{sec8}).

\section{Comparisons between the $T_e$-based (O/H)
and those from empirical strong-line methods}

In this section, we compare the $T_e$-based O/H abundances
of the sample galaxies with
those derived from other ``strong-line" ratios,
such as $R_{23}$, $P$-method and $N$2 (=log([N~{\sc ii}]/H$\alpha$)) index.
The corresponding parameters and the derived (O/H) abundances
are presented in Table~2.

\subsection{ The $R_{23}$ method}

The $R_{23}$ parameter is widely used to estimate metallicities
for metal-rich galaxies.
The empirical relationship between O/H and $R_{23}$
has been suggested and extensively discussed in the literature
(see Pagel et al. 1979; McGaugh 1991; Kobulnicky et al. 1999;  Tremonti et
al. 2004 and references therein; Sect.~\ref{sec4} in this study).
However, some recent observations show that the
 $R_{23}$ will overestimate the log(O/H) abundances by
 0.2-0.5\,dex on  the basis
of about a hundred H~{\sc ii} regions in the spiral galaxies M101 and M51
(KBG03; Bresolin et al. 2004, 2005; Bresolin 2007; Garnett et al. 2004a).

\begin{figure}
\centering
\input epsf
\epsfverbosetrue \epsfxsize 7.8cm
\epsfbox{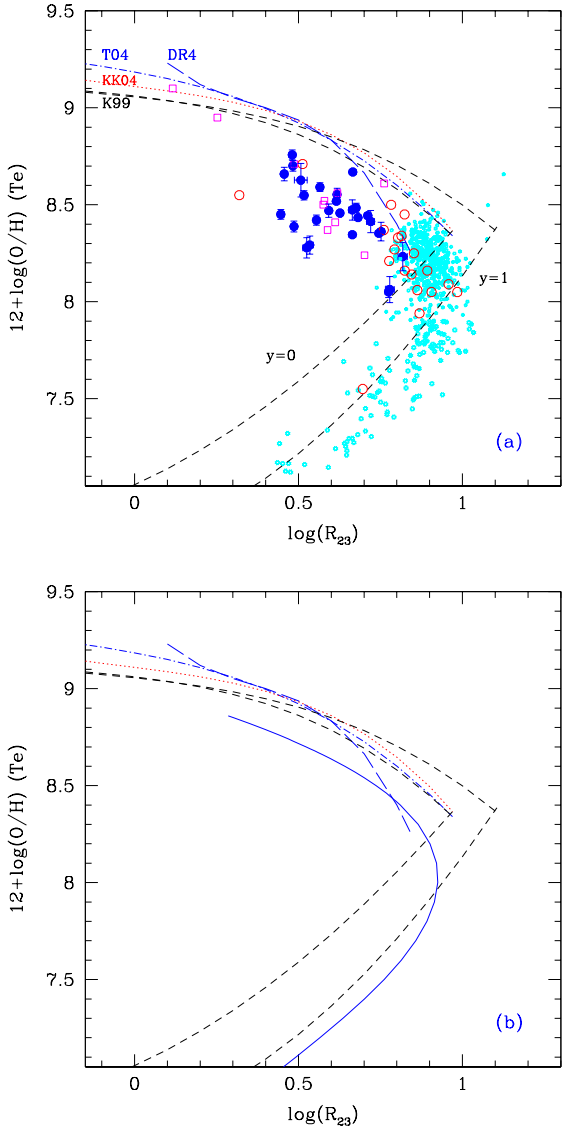}
\caption{{\bf (a)} The relation of the $T_e$-based (O/H) abundances with
$R_{23}$ parameters for our sample galaxies (the filled circles).
Correlation with those from the literature:
 the large open circles refer to the H~{\sc ii} regions in M101 taken from
KBG03, the squares represent the H~{\sc ii} regions given by
CDT02, and the small open circles refer the low-metallicity H~{\sc
ii} regions and galaxies taken from Yin et al. (2007a). Some
empirical calibrations are also given: those of T04 (the
dot-dashed line),
 Kobulnicky et al. (1999; K99, the two dashed-lines,
$y$=log$O_{32}$ = log([O~{\sc iii}]4959,5007/[O~{\sc ii}]3727)),
 Kobulnicky \& Kewley (2004; KK04, the dotted line),
and that derived from SDSS-DR4 in this study (DR4, the long-dashed line).
{\bf (b)} The third-order polynomial fit (the solid line)
for all the observational data points given in (a), also
given as Eq.~(\ref{eqohR23}). 
Other calibrations from the literature are the same as in (a). 
              }
\label{figR23}
\end{figure}

 Figure~\ref{figR23}a shows how these combined SDSS galaxies (the filled circles)
 are distributed in the relationship between log($R_{23}$) and 
 12+log(O/H)$_{T_e}$.
 The empirical $R_{23}$ calibrations obtained by
 T04 (the dot-dashed line),
 K99 (the short-dashed lines), Kobulnicky \& Kewley (2004, KK04; the dotted line)
 and that derived from the SDSS-DR4 database in this study
 (the long-dashed line, see Sect.~\ref{sec4}) are also given.
 The H~{\sc ii} regions studied by KBG03 and Castellanos
 et al. (2002, hereafter CDT02), and metal-poor galaxies 
 studied by Yin et al. (2007a)
 are also plotted. We will not present here other H~{\sc ii} regions 
 from Garnett et al. (2004a) and Bresolin (2007) since these
 are special H~{\sc ii} regions in the center of the spiral galaxies
 and are not exactly the same case as galaxies with their integrated light
 of many H~{\sc ii} regions studied here.

 The combination of all these observational data with $T_e$-based
 (O/H) estimates shows that the $R_{23}$ parameters result in
 double-valued abundances with a transition region of
 7.9-8.2 of 12+log(O/H) and 0.9-1.0 of log($R_{23}$).
 We consistently find that most of our sample galaxies lie in the upper-branch
 with their 8.3$<$12+log(O/H)$<$9.0 and 0.4$<$log($R_{23}$)$<$0.75.
 Our samples are similar to those of CDT02,
 and more metal-rich than those of KBG03 and Yin et al. (2007a).
 Moreover, it is clear that generally the empirical calibrations for upper-branch
 result in higher log(O/H) abundances at a given $R_{23}$, 
 up to 0.6\,dex.

 We obtain a third-order polynomial fit for the total sample of  
 these observational data (702, without considering the 
 two most metal-rich H~{\sc ii} regions from CDT02
 and the three scattered ones with log($R_{23}$)$>$1.0 from the SDSS)
 for log($R_{23}$) vs. log(O/H),
 which is given by the solid
 line in Fig.~\ref{figR23}b, and by
\begin{equation}
  {\rm log}R_{23} = 73.13 - 32.78x + 4.806x^2 -0.230x^3,
 \label{eqohR23}
\end{equation}
 where $x$=12+log(O/H).
 It reveals the discrepancy between this newly derived calibration
 based on the $T_e$-(O/H) abundances and those derived from 
 the previous strong-line empirical calibrations (other lines).
 
\begin{figure}
\centering
\input epsf
\epsfverbosetrue 
\epsfxsize 7.8cm 
\epsfbox{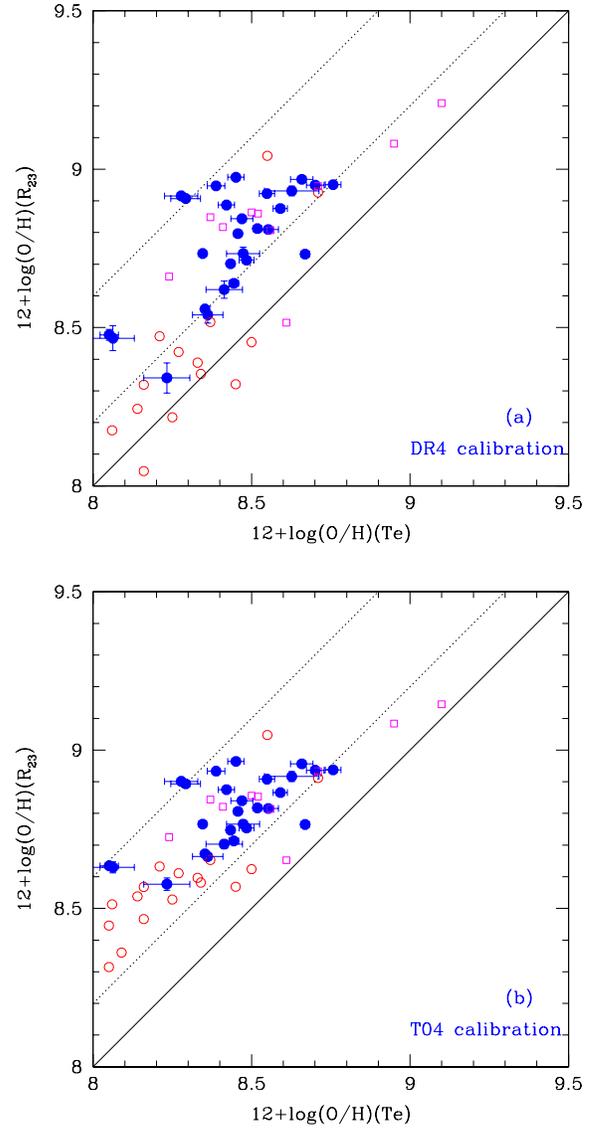}
\caption{Comparison between the $T_e$-based  and  the
$R_{23}$-based log(O/H) abundances of the sample galaxies: {\bf (a)} the
$R_{23}$-calibration was derived from DR4 with 12+log(O/H)$>$8.3 
in this study (Eq.~(\ref{eqOHR23}));
and {\bf (b)}  the $R_{23}$-calibration was derived from DR2 by T04
with 12+log(O/H)$>$8.5. The symbols as the same as in
Fig.~\ref{figR23}.  
The solid line in each panel represents the equal relation, and
the two dashed lines are the relations by increasing 12+log(O/H)$_{R_{23}}$
abundances by 0.2 and 0.6\,dex, respectively.
}
 \label{fig2R23}
\end{figure}

To show this discrepancy more clearly and directly, we compare the
 (O/H)$_{R_{23}}$ and (O/H)$_{T_e}$ in
Fig.~\ref{fig2R23}.
In Fig.~\ref{fig2R23}a, the (O/H)$_{R_{23}}$ abundances of all the data points
are estimated
by using Eq.~(\ref{eqOHR23}), which is the new calibration
 derived from the DR4 database in this study.
In Fig.~\ref{fig2R23}b, the (O/H)$_{R_{23}}$ abundances are
estimated from T04's calibration derived from DR2, which could be
 representative of other empirical calibrations. 
They show that the $R_{23}$ parameter will
overestimate the actual O/H abundances by a factor up to 0.6\,dex
for the moderate metal-rich galaxies. For the objects within the
transition region of abundances, the calibrations of $R_{23}$
derived from DR4 will give consistent abundances as $T_e$, but
others (e.g. DR2) will still overestimate the O/H abundances 
(see the open circles in Fig.~\ref{fig2R23}a,b). 

\subsection{The $P$-method}

 It is worth while comparing the $T_e$-based abundances with those
derived from the $P$-method, which is often assumed to give
``reliable"  metallicities for the metal-rich galaxies (Liang \& Yin 2007).
Pilyugin (2000, 2001a,b; as P01) suggested his
$P$-method to
estimate the oxygen abundances of galaxies in which the oxygen
abundance can be derived from two parameters, $R_{23}$ and $P$
(=[O~{\sc iii}]4959,5007/([O~{\sc ii}]3727+[O~{\sc iii}]4959,5007)). 
 He derived the O/H=$f(R_{23},P)$ formulas from
a sample of metal-poor H~{\sc ii} regions with
7.1$<$12+log(O/H)$_{T_e}$$<$7.95 and a sample of moderately
metal-rich
 H~{\sc ii} regions with 8.2$<$12+log(O/H)$_{T_e}$$<$8.7.  The fitting
relation for moderately metal-rich H~{\sc ii} regions ($\sim$40)
is given as Eq.~(8) in Pilyugin (2001a). Pilyugin \& Thuran (2005;
as PT05) have
renewed the $P$ calibrations by including several
improvements, such as enlarging the sample ($\sim$104 metal-rich
H~{\sc ii} regions). However, on the basis of $\sim$20,000
metal-rich star-forming galaxies from the SDSS-DR2, we found that
the oxygen abundances derived from this revised $P$-method (PT05)
are $\sim$0.19\,dex lower than those derived from the previous one
(P01), and in particular are $\sim$0.60\,dex lower than the Bayesian
abundances obtained by the MPA/JHU group (Yin et al. 2007b). These
abundance discrepancies strongly correlate with the $P$ parameter
and weakly depend on the log($R_{23}$) parameter, which is
especially serious for the case of $P<0.55$ (Yin et al. 2007b;
PT05, P01).

\begin{figure}
\centering
\input epsf
\epsfverbosetrue
\epsfxsize 7.8cm
\epsfbox{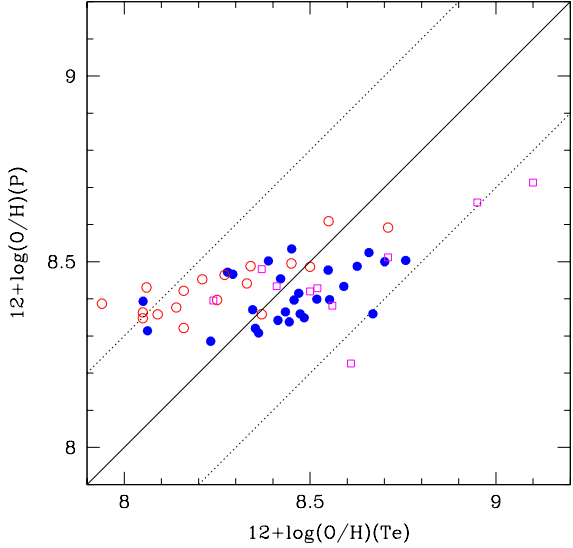}
\caption{Comparison between the $T_e$-based (O/H) abundances and
those estimated from $P$-method for our sample galaxies.
The symbols as the same as in  Fig.~\ref{figR23}.}
\label{fig5}
\end{figure}

In Fig.~\ref{fig5}, we compare the $T_e$-based log(O/H) abundances
of the 27 combined galaxies with those derived by using the
$P$-method of P01 for the metal-rich branch. The H~{\sc ii} regions
studied by KBG03 and CDT02 are also plotted. It seems
that the $P$-method will almost result in a constant (O/H)
abundance of about 12+log(O/H)$\sim$8.4 for both of our samples and
most of the data from the literature, although the log(O/H)$_P$ and log(O/H)$_{Te}$
are consistent within 0.3\,dex. The reason for our sample
galaxies may be that the P01's $P$-method is not quite reliable to
derive their oxygen abundances. This is because all our moderate
metal-rich sample galaxies selected from the SDSS have $P<$0.55
(see Table~2), 
due to the [O~{\sc iii}] line becoming weaker in the metal-rich
environments, 
which is the case for most of the
metal-rich SDSS galaxies (see Yin et al. 2007b). But the
calibration formula of P01 (as well PT05) was derived from fitting
the sample H~{\sc ii} regions that mostly have $P>$0.55, and only a few
of their samples have $P<$0.55. The samples of CDT02 have similar
$P$-values to ours, and show similar trends to those described here. 
The samples of KBG03 generally have higher $P$ values than 0.55.
The reason for their discrepancy from the equal-line may be that
they are just in the turn-over region of abundances when
strong-line calibration cannot give accurate abundances.

\subsection{The $N$2 method}

   The [N {\sc ii}]/H$\alpha$ emission-line ratio
  (as $N2$=log([N {\sc ii}]$\lambda$6583/H$\alpha$)) is useful to
estimate the metallicities of galaxies (Liang et al. 2006; Yin et
al. 2007a, as Y07; Pettini \& Pagel 2004, as PP04; Denicol{\' o}
et al. 2002, as D02; Kewley \& Dopita 2002), though it may depend
on the N-enrichment history of the galaxies (Liang \& Yin 2007),
and ionization parameters (Kewley \& Dopita 2002). $N2$ is not greatly
affected by dust extinction due to the close wavelength positions
of [N {\sc ii}] and H$\alpha$, and the developed near infrared
spectroscopic instruments can gather
 these two lines for the galaxies with intermediate and high redshifts,
 which stand in the early epoch of the universe.

\begin{figure}
\centering
\input epsf
\epsfverbosetrue \epsfxsize 7.8cm 
\epsfbox{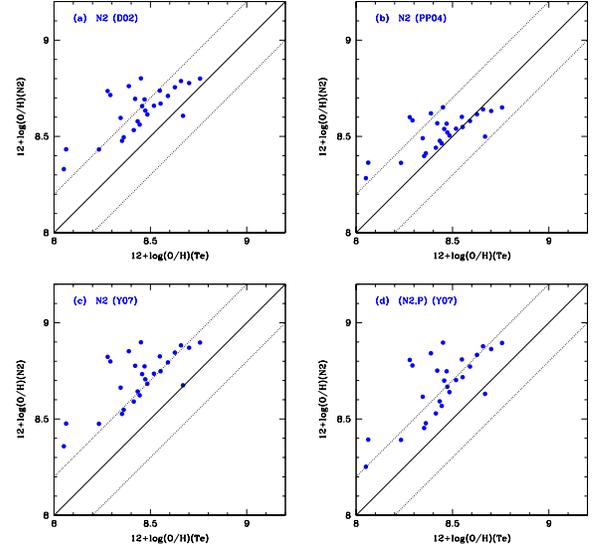}
\caption{Comparisons between the $T_e$-based (O/H) abundances of
the sample galaxies with those estimated from N2 index
calibrations: {\bf (a)} from D02; {\bf (b)} from PP04; {\bf (c)}
from Y07; {\bf (d)} the (N2,P) calibration from Y07.
      }
\label{figN2}
\end{figure}

 Here we compare these
$T_e$-based (O/H) abundances with those derived from the $N2$
calibrations of D02, PP04, Y07, also the $(N2,P)$ calibration from Y07 ($P$
=[O~{\sc iii}]/([O~{\sc ii}]+[O~{\sc iii}])), given in
Fig.~\ref{figN2}a-d. They show that the $N2$-calibration given by
PP04 could give (O/H) abundances that are more consistent with the
$T_e$-based ones, while other
$N2$-calibrations often overestimate the (O/H) abundances for
these moderate metal-rich galaxies.

\section{The log(N/O) abundance ratios}

   The ``direct" electron temperature in the low-ionization region
   of the galactic gas and the high quality optical spectra
   make it possible to estimate the log(N/O) abundances
   of these combined galaxies properly, which is useful to understand the
   ``primary" and/or ``secondary" origin of the nitrogen element.

     If the ``seed" oxygen
 and carbon for the production of nitrogen via the CNO processing
 are those incorporated into a star at its formation and a
 constant mass fraction is processed, then the amount of nitrogen produced
 is proportional to the initial heavy-element abundance, and the nitrogen
 synthesis is said to be ``secondary". If the oxygen and carbon are produced
 in the star prior to the CNO cycling (e.g. by helium burning in a core,
 followed by CNO cycling of this material mixed into a hydrogen-burning
 shell), then the amount of nitrogen produced may be fairly independent of
 the initial heavy-element abundance of the star, and the synthesis is said
 to be  ``primary" (Vila-Costas \& Edmunds 1993). In general, primary
 nitrogen production is independent of metallicity, while secondary
 production is a linear function of it.

\begin{figure}
\centering
\input epsf
\epsfverbosetrue
\epsfxsize 7.8cm
\epsfbox{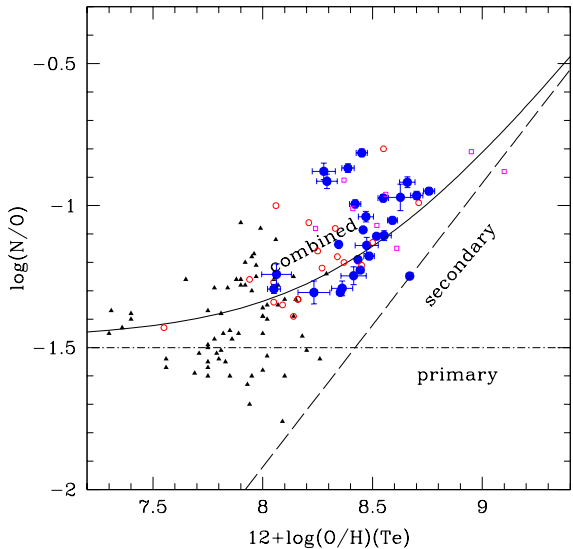}
\caption{Log(N/O) abundance ratios of the sample galaxies
 as functions of their 12+log(O/H) abundances derived from $T_{e}$
 (the large filled circles).
  The dot-dashed line refers to the ``primary" component of nitrogen,
  the long-dashed line refers to the ``secondary" component of nitrogen,
  and the solid line refers to the combination of these two
 components, which are taken from Vila-Costas \& Edmunds (1993). Other recent
 observations are also given: the triangles represent the dwarf galaxies
 studied by van Zee \& Haynes (2006), and
 the open circles and squares represent the extragalactic H~{\sc ii}
 regions studied by KBG03 and CDT02, respectively. }
\label{fig7}
\end{figure}

   We adopt the low-ionization region temperature $t_2$ for [N~{\sc ii}]
   emission regions
   ($t_2$([N~{\sc ii}])= $t_2$([O~{\sc ii}])) and the
    [N~{\sc ii}]$\lambda\lambda$6548,6583/[O~{\sc ii}]$\lambda$3727 ratio
  to derive the log(N/O) abundances of these combined galaxies
  by using the formulas of Izotov et al. (2006) for the 
  log($\rm {N^+}\over{H^{+}}$) and log($\rm {O^+}\over{H^{+}}$) abundances,
  with $\rm {N}\over{O}$=$\rm {N^+}\over{O^{+}}$.
  Figure~\ref{fig7} presents the results.
  These $T_e$-based (O/H) and (N/O) abundances show that
  the sources of nitrogen for these moderate metal-rich galaxies are
  consistent with the combination of the primary and
  secondary components.
  This is a robust investigation on the
  production of nitrogen for these moderate metal-rich galaxies
  since both the N and O abundances are obtained from
  electron temperature on the basis of
  their integrated light.
  The H~{\sc ii} regions of KBG03 and CDT02 also show
  a similar nitrogen source to our samples.
  Most of the dwarf galaxies taken from
  van Zee \& Haynes (2006) show the primary source of their nitrogen.

\section{Discussion}
\label{sec8}

Our procedure to derive $T_e$-based oxygen abundances for the metal-rich
galaxies may be affected by some biases that need to be discussed further.
The major contributors for such biases are the dust estimates, the
stacking method, and the uncertainties related to the temperature
determination.

\subsection{Dust extinction}

Our method is strongly dependent on the dust extinction, since the
determination of the temperature depends on the ratio of red over blue emission
lines. However, the S/N of the stacked spectra is so large that our estimate of
$A_V$ is very accurate. On the other hand, one may wonder whether by stacking
spectra, we still present the representative of the mean properties of individual
galaxies.

\begin{figure}
\centering
\input epsf
\epsfverbosetrue
\epsfxsize 7.8cm
\epsfbox{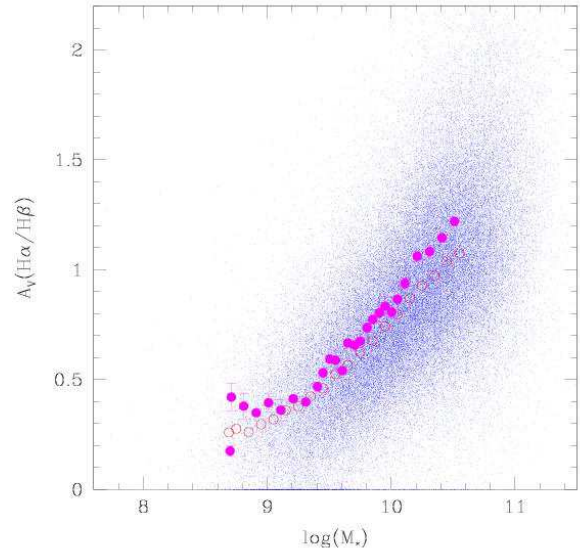}
\caption{Relationship between dust extinction and stellar masses of the sample
galaxies. The large filled circles refer to the results from the stacked spectra 
of the galaxies in the 27 mass bins.  
The small points
refer to the star-forming galaxies from SDSS DR4 
(the same as in Fig.~\ref{figdr4}a),
and the open circles refer to the median values of them.
(Please see the online color version of the plot for the points,
especially for the open circles.)
 }
\label{fig12}
\end{figure}

We confirm the strong correlation between the dust extinction ($A_V$) and the
stellar masses of star-forming galaxies, which, as expected, means the more
massive galaxies generally have more heavy dust attenuation (see also Salim et
al. 2007). In Fig.~\ref{fig12} the large filled circles represent our stacked
galaxies, the small points represent the whole
sample of the star-forming galaxies from 
SDSS DR4 (same as in Fig.~\ref{figdr4}a),
and the large open circles 
represent the median values of this whole sample in the corresponding mass bins. 
In most of the considered
stellar mass range, the stacked spectra provide a $A_V$-$M_*$ 
relationship almost
identical to that of the whole sample, meaning that the stacking method and
selection criteria do not cause significant biases in 
the estimation of the dust
extinction. Notice however that at the highest mass range, stacking spectra
provide a slightly higher extinction, which may well be due to the selection
method. Indeed, by selecting stronger star forming galaxies than the average,
it is not surprising to find slightly higher dust extinction.
   
\subsection{Are stacked spectra providing a good representation of real galaxies ?}

 The ``stacking" method has been used widely in astronomical study,
 e.g., Zheng et al. (2006) and Chang et al. (2006), 
 who stacked the images or property parameters of galaxies
 to study their properties.
  There are also examples of uses of stacked spectra:
   Mathis et al. (2006) stacked the spectra
   in several mass bins to study the cosmic star formation history;
   Baldry et al. (2002) analyzed the stacked
   low-resolution spectra of 166,000 galaxies at redshift
   0.03$<$$z$$<$0.25 extracted from the 2dFGS;
   Schiavon et al. (2006) used the stacked Keck DEIMOS
   spectra to study the ages and metallicities of some red galaxies;
   and Wild et al. (2007) used the stacked SDSS
   spectra in redshift bins to study the [O~{\sc ii}]$\lambda$3727 nebular 
   emission from the
   galaxies hosting Ca~{\sc ii} absorption line systems and galaxies 
   hosting Mg~{\sc ii}-selected DLA absorbers.
 
   We are however conscious that the stacking method may produce undesirable
 biases since it is unclear how some physical properties may be affected by
 this method. In Sect.~\ref{sec4} and Sect.~\ref{sec8}.1, we have
 convincingly shown that the stacking method does not cause significant biases
 in representing the median value for the O/H estimated values from the
 $R_{23}$ method, and for the extinction estimates, respectively. Both
 estimates are based on ratios of lines at different wavelengths. It is thus
 tempting to believe that temperatures, estimated in a similar way, would not
 be affected by the stacking method. However, we agree that this needs a full
 demonstration, which is beyond the scope of this paper. We also notice that
 the integrated spectrum of a single galaxy comes from all the contributed
 star-forming regions. By using stacked spectra
 from different galaxies, we are studying the resulted light coming from a
 larger set of star-forming regions. 
 In principle our method is not less
 accurate than the determination of the O/H through temperature for a single
 galaxy, and thus may not alter the calibrations proposed in this paper.

 The major interest of the adopted stacked method presented here, is that it
allows us for the first time to sample the  stellar mass versus log(O/H) relation
using a temperature dependent method, up to relatively high stellar mass
galaxies.

\subsection{The uncertainties of the T([O II]$\lambda$7325)-based abundances}

In the optical band, electron temperatures could usually be
derived using five sets of auroral/nebular line intensity ratios:
[O~{\sc iii}]$\lambda$4363/[O~{\sc iii}]$\lambda$$\lambda$4959,
5007, [S~{\sc iii}]$\lambda$6312/ [S~{\sc iii}]$\lambda$$\lambda$9069, 9532, 
[S~{\sc ii}]$\lambda$$\lambda$6317,6731/[S~{\sc ii}]$\lambda$4072,
[N~{\sc ii}]$\lambda$5755/[N~{\sc ii}]$\lambda$$\lambda$6548,6583, and 
[O~{\sc ii}]$\lambda$$\lambda$7320,7330/ [O~{\sc ii}]$\lambda$3727.
However, for the integrated spectra of the SDSS sample galaxies
with moderate stellar masses, only the  indicator [O~{\sc
ii}]$\lambda$$\lambda$7320,7330/[O~{\sc ii}]$\lambda$3727 is
available to derive their electron temperatures in the galactic
gas since only [O~{\sc ii}]$\lambda$$\lambda$7320,7330 lines
(marked as $\lambda$7325 sometimes) are strong enough to be
measured from these spectra. Therefore, the line ratio
[O~{\sc ii}]$\lambda$$\lambda$7320,7330/[O~{\sc ii}]$\lambda$3727 provides
the only ``direct" measurements for the electron temperature and
the ``direct" oxygen abundances for these moderate metal-rich
galaxies, though some arguments exist to doubt the accuracy
discussed as following.

One of the problems of $T(7325)$ is the contribution of recombination
to the  [O~{\sc ii}]$\lambda$7325 line. However,
this has been estimated by 
KBG03 using the formula given by Liu et al. (2000),
who found that recombination typically contributes less than 5\% to the
[O~{\sc ii}] line flux, which corresponds to a temperature error of only
$\sim$2\%-3\% or less than 400K for their worst case.
A correct treatment of recombination
should take into account the effect of temperature gradients within ionized
nebulae.

The second problem is the inconsistency between the $T(7325)$ and the
 $T(5755)$ and $T(4072)$, presented by Bresolin et al. (2005).
By using the VLT/FORS spectra, Bresolin et al. (2005) obtained the
$T$(7325) for 32 H~{\sc ii} regions located in some spiral
galaxies, and $T(5755)$ and $T(6312)$  for only half of
their sample since these two lines are much weaker. In addition,
[S~{\sc ii}]4072, from which $T(4072)$ was derived, has been
seldom detected, since its measurement is made difficult by low S/N
ratio in the spectra. They compared these temperature values, and
found that $T(7325)$ seems to overestimate the temperature if
compared to $T(5755)$, while the opposite happens for $T(4072)$.
However, the error-bars of the latter two temperatures are much
larger than $T(7325)$, and these two lines are much weaker than
$\lambda$7325. We cannot measure the other temperature-sensitive
auroral lines from most of our sample spectra.

Another problem is that the $T$\oii~ vs. $T$\oiii~ relation does
not match the model fitting relation very well (see Fig.~1 of
 KBG03; Garnett 1992). They mentioned that
the sources of this disagreement remain unresolved,
but one of the sources may be the
recombination contribution (Liu et al. 2000),
others could be the radiative transfer effects and the observational
uncertainties.
However, their objects with large scatter mostly have
 $T$\oiii~ between 9000K and 12000K, and there may
be a correlation between $T$\oii~ and $T$\oiii~ in the extremes of the 
temperature range, e.g. 
$T_e<$9000K and $T_e>$13000K. Most of our sample galaxies
have just $T$\oiii$<$8000K, thus no direct observations are
against that our galaxies follow
that correlation of $T$\oii~ vs. $T$\oiii.

Therefore, the temperature derived from [O~{\sc
ii}]$\lambda$7325 is acceptable, and is the only way to get the
``direct" oxygen abundances for metal-rich
galaxies.

\subsection{Temperature fluctuations and temperature gradients}

It has been argued for many years that directly measured electron temperatures
and the corresponding abundances from collisionally excited lines may have
systematic errors due to temperature fluctuations and temperature gradients,
especially for the metal-rich H~{\sc ii} regions
(Peimbert 1967; Stasinska 1980; Stasinska 2005).
However, the auroral lines have been used to derive the ``direct"
electron temperature and the oxygen abundances for the metal-rich
 H~{\sc ii} regions in recent years (KBG03; Garnett et al. 2004a,b;
 Bresolin et al. 2004,2005; Bresolin 2007).
 KBG03 has discussed very clearly the ``reliability" of the auroral line
 method for abundance estimates. We simply summarize some arguments and
 discussions
 presented by KBG03, Garnett (1992) and Stasinska (2005) regarding this topic.

In the presence of fluctuations in $T_e$, forbidden line strengths can
overestimate the average temperature, and therefore
underestimate the true nebular abundances.
However, the question of the existence of temperature fluctuations is still
controversial (Peimbert 1967; Dinerstein et al. 1985; Shaver et al. 1983).
The
large-scale temperature gradient mimics the effects of temperature fluctuations,
 and can over-estimates the real temperature, and consequently the
oxygen abundance is systematically under-estimated
(Garnett 1992; Stasinska 2005).

 Despite these concerns, as KBG03 discussed,
 there are good reasons to believe that
 the forbidden lines give abundances that are close to correct,
 at least for the range of O/H of interest here. Recent measurements of
 infrared forbidden lines (that have weak temperature dependences)
 in the spectra of planetary nebulae yield abundances in good
 agreement with those from optical forbidden lines,
 even where there is a large discrepancy with recombination line abundances
 (Liu et al. 2000, 2001). Likewise, measurements of radio
 recombination line temperatures of Galactic H~{\sc ii} regions give results that
 are consistent with forbidden line measurements of the same objects
 (Shaver et al. 1983; Deharveng et al. 2000).
In both cases, the line emissivity is only
moderately dependent on $T_e$.
It would be worthwhile to explore the
metal recombination lines and infrared fine-structure lines
for the determination of electron temperature in the 
extragalactic \hii\/ regions
of high metal content. However, there are obvious difficulties in
exploiting these techniques: recombination lines are very weak
and hard to detect in extragalactic \hii\/ regions
(Esteban 2002), and infrared observations require orbiting
telescopes (Garnett et al. 2004b).

\section{Conclusion}

We have derived direct oxygen abundances based on electron
temperature for metal-rich galaxies, using an original method of
stacking. To do so, we select a large sample of star-forming
galaxies ($\sim$23,608) from the SDSS-DR4 and obtain 27 spectra by
combining the spectra of the several hundred (even several thousand)
galaxies within 27 various stellar mass bins from log($M_*$)=8.0
to 10.6. This stacking method improves the S/N ratios in a great deal of the
spectra, without providing major biases. These
high quality data with moderate resolution (3\AA) make it
possible for us to measure the [O~{\sc
ii}]$\lambda$$\lambda$7320,7330 emission-lines for the galaxies
with stellar masses up to log($M_*$)$\sim$10.6, then derive their
electron temperature in the low ionization region by the ratio of
[O~{\sc ii}]$\lambda$$\lambda$7320,7330/[O~{\sc ii}]$\lambda$3727,
hence the $T_e$-based O/H abundances.

 It provides a new calibration of the $R_{23}$ method, which
may well apply to galaxies with metallicities ranging from
12+log(O/H)$\sim$7.0 to 9.0. It also shows that the empirical
$R_{23}$ calibrations overestimate the log(O/H) abundances by
about 0.2 to 0.6\,dex. We also derive a new relationship between
 $T_e$-derived (O/H) abundances and stellar masses from these 
 moderate massive galaxies,
 which can be fitted by a linear fit, namely
 12+log(O/H) =6.223+0.231$\times$log($M_*$). 
The $T_e$-based log(N/O) abundance ratios show that their nitrogen
sources are consistent with the combination of the primary
and secondary components for these moderate metal-rich galaxies.

\section*{Acknowledgments}
  We especially thank our referee for the very valuable
  comments and suggestions, which helped greatly in improving this work. 
  We thank Jing Wang and Grazyna Stasinska for helpful discussions about the
  underlying correction of stellar absorption for the SDSS spectra,
  thank Hong Wu, Jingyao Hu, Licai Deng and Bo Zhang for interesting discussions.
  This work was supported by the Natural Science Foundation of China
 (NSFC) Foundation under No.10403006, 10433010, 10673002, 
 10573022, 10333060, and 10521001; and the
 National Basic Research Program of China (973 Program) No.2007CB815404.
 The STARLIGHT project is supported by the Brazilian agencies CNPq, 
 CAPES and FAPESP and by the France-Brizil CAPES/Cofecub program.


\begin{table*} [b]
\begin{center}
\caption{\footnotesize The derived properties of the combined galaxies.}
\label{tab2}
{\tiny
\begin{tabular}{llllllr|llllllllll}
\hline \hline
(1)  & (2) & (3) & (4) & (5) & (6) & (7) & (8)  & (9) & (10) & (11) & (12) & (13) & (14) & (15) & (16) & (17)\\ \hline
   mass   & s2 ratio  &$n_e$      & o2 ratio  & $T_e$([O~{\sc ii}])  &
   $T_e$([O~{\sc iii}])  & OH$_{T_e}$ & $A_V$ & log$R_{23}$ & log$O_{32}$ &
   OH$^a_{R_{23}}$ &OH$^b_{R_{23}}$ & $P$ & OH$_{P}$  & N2 & OH$_{N2}$ &  log(${\rm N}\over {\rm O}$)    \\ 
       &    &  (cm$^{-3}$)    &    & (K) &
     (K) &   &   &   &   &
     &  &    &  &   &       \\ \hline
    8.70    &   1.36   &    58.97    &    44.25        &   11423       &   12034      &   8.05          &   0.18   &  0.77  &    0.08  &   8.48   &   8.63  &  0.54 &  8.39  & -1.08  &    8.28   &   -1.29   \\
            &          &             &    $\pm$   0.95  &  $\pm$   153  &  $\pm$   219  &   $\pm$  0.03   &          &    & 	      &       &      &    &     &    &       &       \\
    8.71    &   1.38   &    39.36    &    51.32        &   10641       &   10916      &   8.23          &   0.42   &  0.82  &   -0.05  &   8.34   &   8.58  &  0.47 &  8.29  & -0.94  &    8.36   &   -1.31   \\
            &          &             &    $\pm$   2.76  &  $\pm$   341  &  $\pm$   487  &  $\pm$  0.07    &           &    & 	      &       &      &    &     &    &       &       \\
    8.81    &   1.43   &     3.58    &    47.38        &   11491       &   12130       &   8.06         &   0.38   &  0.78  &   -0.11  &   8.47   &   8.63  &  0.44 &  8.31  & -0.94  &    8.36   &   -1.24   \\
            &          &             &    $\pm$   2.39  &  $\pm$   370  &   $\pm$   528  &   $\pm$  0.07   &            &    & 	      &       &      &    &     &    &       &       \\
    8.91    &   1.41   &    14.06    &    63.38        &   9703        &    9576       &   8.35         &   0.35   &  0.75  &   -0.21  &   8.56   &   8.67  &  0.38 &  8.32  & -0.88  &    8.40   &   -1.31   \\
            &          &             &     $\pm$   0.29 &   $\pm$    22  &   $\pm$   32  &  $\pm$  0.01   &            &    & 	      &       &      &    &     &    &       &       \\
    9.01    &   1.40   &    24.83    &    62.37         &   9701       &    9572       &   8.36         &   0.39   &  0.75  &   -0.22  &   8.54   &   8.66  &  0.38 &  8.31  & -0.86  &    8.41   &   -1.29   \\
            &          &             &     $\pm$   2.23  & $\pm$   182  &  $\pm$   260   &   $\pm$  0.05   &           &    & 	      &       &      &    &     &    &       &       \\
    9.11    &   1.37   &    50.84    &    64.91         &   9328       &    9039       &   8.41          &   0.36   &  0.72  &   -0.24  &   8.62   &   8.70  &  0.36 &  8.34  & -0.80  &    8.44   &   -1.25   \\
            &          &             &    $\pm$   2.75  &  $\pm$   202  &   $\pm$   289  &    $\pm$  0.06  &           &    & 	      &       &      &    &     &    &       &       \\
    9.21    &   1.39   &    33.32    &    69.04         &   9168       &    8812       &   8.44         &   0.41   &  0.71  &   -0.28  &   8.64   &   8.71  &  0.34 &  8.34  & -0.77  &    8.46   &   -1.23   \\
            &          &             &    $\pm$   0.80  &  $\pm$    51  &   $\pm$    72  &  $\pm$  0.02   &            &    & 	      &       &      &    &     &    &       &       \\
    9.31    &   1.38   &    40.24    &    69.45         &   9096       &    8709       &   8.43         &   0.40   &  0.68  &   -0.31  &   8.70   &   8.75  &  0.33 &  8.36  & -0.74  &    8.48   &   -1.19   \\
            &          &             &     $\pm$   0.39  & $\pm$    24  &   $\pm$    34  &   $\pm$  0.01   &           &    & 	      &       &      &    &     &    &       &       \\
    9.405   &   1.36   &    57.15    &    63.03         &   9422       &    9174       &   8.35          &   0.47   &  0.66  &   -0.36  &   8.73   &   8.77  &  0.30 &  8.37  & -0.72  &    8.49   &   -1.14   \\
            &          &             &     $\pm$   0.50  & $\pm$    38  &   $\pm$    54  &   $\pm$  0.01   &           &    & 	      &       &      &    &     &    &       &       \\
    9.4505  &   1.39   &    37.63    &    73.34         &   8883       &    8404       &   8.48         &   0.53   &  0.68  &   -0.39  &   8.71   &   8.75  &  0.29 &  8.35  & -0.69  &    8.50   &   -1.18   \\
            &          &             &     $\pm$   1.29  & $\pm$    74  &   $\pm$   105  &  $\pm$  0.02   &            &    & 	      &       &      &    &     &    &       &       \\
    9.505   &   1.42   &    13.25    &    92.13         &   8169       &    7384       &   8.67        &   0.59   &  0.67  &   -0.39  &   8.73   &   8.77  &  0.29 &  8.36  & -0.70  &    8.50   &   -1.25   \\
            &          &             &     $\pm$   0.91  &  $\pm$    34 &   $\pm$    49  &  $\pm$  0.01   &           &    & 	      &       &      &    &     &    &       &       \\
    9.5505  &   1.42   &    10.81    &    76.59         &   8878       &    8397       &   8.47        &   0.59   &  0.66  &   -0.40  &   8.73   &   8.77  &  0.29 &  8.36  & -0.67  &    8.52   &   -1.14   \\
            &          &             &     $\pm$   2.96  & $\pm$   164  &  $\pm$   235   & $\pm$  0.05   &            &    & 	      &       &      &    &     &    &       &       \\
    9.605   &   1.40   &    29.90    &    75.72         &   8801       &    8287       &   8.46        &   0.54   &  0.63  &   -0.42  &   8.80   &   8.81  &  0.27 &  8.40  & -0.63  &    8.54   &   -1.09   \\
            &          &             &     $\pm$   0.09  & $\pm$     5  &  $\pm$     7   & $\pm$  0.00   &            &    & 	      &       &      &    &     &    &       &       \\
    9.6505  &   1.38   &    41.11    &    79.67         &   8534       &    7906       &   8.52        &   0.67   &  0.62  &   -0.46  &   8.81   &   8.82  &  0.26 &  8.40  & -0.63  &    8.54   &   -1.11   \\
            &          &             &     $\pm$   0.77  & $\pm$    37  &   $\pm$    53  & $\pm$  0.01   &            &    & 	      &       &      &    &     &    &       &       \\
    9.705   &   1.39   &    32.46    &    83.36         &   8414       &    7735       &   8.55        &   0.66   &  0.62  &   -0.45  &   8.81   &   8.82  &  0.26 &  8.40  & -0.62  &    8.55   &   -1.11   \\
            &          &             &     $\pm$   1.99  & $\pm$    90  &   $\pm$   128  &  $\pm$  0.03   &           &    & 	      &       &      &    &     &    &       &       \\
    9.7505  &   1.38   &    39.36    &    77.80         &   8636       &    8051       &   8.47        &   0.68   &  0.59  &   -0.50  &   8.84   &   8.84  &  0.24 &  8.42  & -0.59  &    8.57   &   -1.04   \\
            &          &             &    $\pm$   1.99   & $\pm$   102  &   $\pm$   146  &  $\pm$  0.03   &            &    & 	      &       &      &    &     &    &       &       \\
    9.805   &   1.38   &    39.36    &    76.95         &   8679       &    8112       &   8.42         &   0.74   &  0.55  &   -0.50  &   8.89   &   8.88  &  0.24 &  8.45  & -0.58  &    8.57   &   -0.99   \\
            &          &             &     $\pm$   1.50  &  $\pm$    75  &  $\pm$   108  &  $\pm$  0.03   &            &    & 	      &       &      &    &     &    &       &       \\
    9.8505  &   1.38   &    42.86    &    89.25         &   8114        &    7305      &   8.59         &   0.77   &  0.57  &   -0.54  &   8.88   &   8.87  &  0.22 &  8.43  & -0.56  &    8.58   &   -1.05   \\
            &          &             &     $\pm$   1.48  &  $\pm$    58  &  $\pm$    83  &  $\pm$  0.02   &             &    & 	      &       &      &    &     &    &       &       \\
    9.905   &   1.39   &    32.46    &    69.83         &   9124        &    8748      &   8.29        &   0.80   &  0.53  &   -0.54  &   8.91   &   8.89  &  0.23 &  8.47  & -0.56  &    8.58   &   -0.91   \\
            &          &             &    $\pm$   2.42   & $\pm$   156  &   $\pm$   224  & $\pm$  0.05   &             &    & 	      &       &      &    &     &    &       &       \\
    9.9505  &   1.36   &    64.48    &    66.14          &   9156      &    8794       &   8.28        &   0.83   &  0.53  &   -0.55  &   8.92   &   8.90  &  0.22 &  8.47  & -0.53  &    8.60   &   -0.88   \\
            &          &             &     $\pm$   2.55  &  $\pm$   177  &  $\pm$   253  & $\pm$  0.05   &             &    & 	      &       &      &    &     &    &       &       \\
   10.005   &   1.37   &    54.43    &    88.12         &   8096        &    7281      &   8.55        &   0.81   &  0.52  &   -0.56  &   8.92   &   8.91  &  0.21 &  8.48  & -0.52  &    8.60   &   -0.97   \\
            &          &             &     $\pm$   1.56  &  $\pm$    62  &  $\pm$    88  &  $\pm$  0.02   &             &    & 	      &       &      &    &     &    &       &       \\
   10.0505  &   1.38   &    40.24    &    98.15         &   7813        &    6876      &   8.63         &   0.87   &  0.51  &   -0.56  &   8.93   &   8.92  &  0.22 &  8.49  & -0.50  &    8.61   &   -0.97   \\
            &          &             &     $\pm$   6.10  &   $\pm$   210  & $\pm$   299  &  $\pm$  0.09   &             &    & 	      &       &      &    &     &    &       &       \\
   10.11    &   1.34   &    76.67    &    76.26         &   8563        &    7947      &   8.39        &   0.94   &  0.49  &   -0.58  &   8.95   &   8.93  &  0.21 &  8.50  & -0.49  &    8.62   &   -0.87   \\
            &          &             &     $\pm$   1.60  &   $\pm$    82  & $\pm$   118  &  $\pm$  0.03   &             &    & 	      &       &      &    &     &    &       &       \\
   10.21    &   1.34   &    75.72    &   101.92         &   7531        &    6472      &   8.70        &   1.06   &  0.48  &   -0.61  &   8.95   &   8.94  &  0.20 &  8.50  & -0.47  &    8.63   &   -0.96   \\
            &          &             &    $\pm$   2.07  &   $\pm$    61  &  $\pm$    88  & $\pm$  0.03   &             &    & 	      &       &      &    &     &    &       &       \\
   10.31    &   1.32   &    97.12    &    97.23        &   7579         &    6542      &   8.66        &   1.08   &  0.46  &   -0.61  &   8.97   &   8.96  &  0.20 &  8.52  & -0.46  &    8.64   &   -0.92   \\
            &          &             &    $\pm$   2.54  &   $\pm$    77  &  $\pm$   110  &  $\pm$  0.04   &             &    & 	      &       &      &    &     &    &       &       \\
   10.41    &   1.33   &    86.29    &    81.74        &   8197         &    7424      &   8.45        &   1.15   &  0.45  &   -0.60  &   8.97   &   8.96  &  0.20 &  8.53  & -0.44  &    8.65   &   -0.81   \\
            &          &             &     $\pm$   1.54  &  $\pm$    68  &  $\pm$    97  &  $\pm$  0.03   &             &    & 	      &       &      &    &     &    &       &       \\
   10.51    &   1.36   &    57.15    &   110.86         &   7368        &    6241      &   8.76        &   1.22   &  0.48  &   -0.60  &   8.95   &   8.94  &  0.20 &  8.50  & -0.44  &    8.65   &   -0.95   \\ 
            &          &             &    $\pm$   1.85  &   $\pm$    48  &  $\pm$    69  &  $\pm$  0.02   &           &    & 	      &       &      &    &     &    &       &        \\ \hline
\end{tabular}  
} 
\end{center}
{\scriptsize Notes: Col.~(1)-(17) consequently refer to:
(1) the mass of the center value of the bin;
(2) the ratio of [S~{\sc ii}]$\lambda6717$/[S~{\sc ii}]$\lambda6731$;
(3) the electron density;
(4) the ratio of [O~{\sc ii}]$\lambda\lambda7320,7330$/[O~{\sc ii}]$\lambda3727$;
(5) the election temperature in low ionization region;
(6) the election temperature in high ionization region;
(7) the oxygen abundances 12+log(O/H) derived from election temprature;
(8) the dust extinction $A_V$ estimated from H$\alpha$/H$\beta$;
(9) logarithm of the ratio of ([O~{\sc ii}]$\lambda3727$+[O~{\sc
iii}]$\lambda\lambda4959,5007$)/H$\beta$;
(10) logarithm of the ratio of [O~{\sc iii}]$\lambda\lambda4959,5007$/[O~{\sc ii}]$\lambda3727$;
(11) the 12+log(O/H) abundances obtained by using
the $R_{23}$ calibrations derived from DR4 (Eq.~\ref{eqOHR23});
(12) the 12+log(O/H) abundances obtained by using
the $R_{23}$ calibrations derived from DR2 (T04);
(13) the $P$ parameter [O~{\sc iii}]$\lambda\lambda4959,5007$)/
([O~{\sc ii}]$\lambda3727$+[O~{\sc iii}]$\lambda\lambda4959,5007$);
(14) the 12+log(O/H) abundances obtained by using the $P$-method calibration of P01;
(15) the $N2$ index (=log([N~{\sc ii}]$\lambda6583$/H$\alpha$));
(16) the 12+log(O/H) abundances obtained by using the $N2$ calibration of PP04;
(17) the log(N/O) abundance ratios derived from election temperature method. 
}
\end{table*}                                      


\begin{thebibliography}{}

\bibitem[2007] {A07} Asari, N. V. et al. 2007, MNRAS, submitted

\bibitem[2002] {B02} Baldry, I. K. et al., 2002, ApJ, 569, 582

\bibitem[2004] {B04} Bresolin, F., Garnett, D.R., \& Kennicutt, R. C. Jr., 2004, ApJ, 615, 228

\bibitem[2005] {B05} Bresolin, F., Schaerer, D. et al. 2005, A\&A, 441, 981

\bibitem[2006] {B07} Bresolin, F. 2007, ApJ, astro-ph/0610690

\bibitem[2004]{Br04} Brinchmann, J., Charlot, S., White, S.~D.~M.,
Tremonti, C., Kauffmann, G., Heckman, T. \& Brinkmann, J., 2004, MNRAS, 351, 1151

\bibitem[2003] {B03} Bruzual, A.G., Charlot, S., 2003, MNRAS, 344, 1000 (BC03)

\bibitem[1989] {C89} Cardelli, J. A., Clayton, G. C., Mathis, J. S., 1989, ApJ, 345, 245

\bibitem[2002] {C02} Castellanos, M., Díaz, A. I., Terlevich, E., 2002, MNRAS, 337, 540 (CDT02)

\bibitem[2003] {C03}  Chabrier, G., 2003, PASP, 115, 763

\bibitem[2001] {C01} Charlot, S. \& Longhetti, M., 2001, MNRAS, 323, 887  (CL01)

\bibitem[2006] {C06} Chang, R. X., Shen, S. Y., Hou, J. L., Shu, C. G., Shao, Z. Y.,
2006, MNRAS, 372, 199

\bibitem[2005] {C05}  Cid Fernandes, R., Mateus, A., Sodre, L. et al. 2005, MNRAS, 358, 363

\bibitem[2007] {C07} Cid Fernandes, R. et al. 2007, RevMexAA, in preperation

\bibitem[1987] {d87} de Robertis, M.M., Dufour, R. J., Hunt, R.W., 1987, JRASC, 81, 195

\bibitem[2002] {D02} Denicol{\' o}, G., Terlevich, R., Terlevich, E., 2002, MNRAS, 330, 69 (D02)

\bibitem[1985] {D85} Dinerstein, H.L., Lester, D.F., Werner, M.W., 1985, ApJ, 291, 561

\bibitem[2002] {E02} Esteban, C., Peimbert, M., Torres-Peimbert, S., Rodríguez, M., 2002,
ApJ, 581, 241

\bibitem[1992] {G92} Garnett, D.R., 1992, AJ, 103, 1330

\bibitem[2004] {G04a} Garnett, D.R., Kennicutt, R.C. Jr., Bresolin, F., 2004a, ApJ, 607, L21

\bibitem[2004] {G04b} Garnett, D.R., Edmunds, M.G., Henry, R.B.C., Pagel, B.E.J., Skillman, E.D., 2004b, AJ, 128, 2722

\bibitem[2006] {I06} Izotov, Y.I., Stasinska, G., Meynet, G., Guseva, N.G., Thuan, T.X.,
2006, A\&A, 448, 955
\bibitem[2003] {K03} Kauffmann, G., Heckman, T.M., Tremonti, C.A., et al. 2003, MNRAS, 346, 1055

\bibitem[2003] {K03} Kennicutt, R.C.Jr., Bresolin, F., Garnett, D.R., 2003, ApJ, 591, 801
(KBG03)

\bibitem[2002] {K02} Kewley, L.J., \& Dopita, M.A., 2002, ApJS, 142, 35

\bibitem[1999] {K99} Kobulnicky, H.A., Kennicutt, R.C.Jr., \& Pizagno, J.L., 1999, ApJ, 514, 544 (K99)

\bibitem[2004] {K04} Kobulnicky, H.A., Kewley, L. J, 2004, ApJ, 617, 240 (KK04)

\bibitem[2006] {L06} Liang, Y.C., Yin, S.Y., Hammer, F. et al. 2006, ApJ, 652, 257

\bibitem[2006] {L06} Liang, Y.C. \& Yin, S.Y., 2007 ,proceedings of "The central
engine of Active Galactic Nuclei" eds. L. C. Ho and J.-M. Wang, astro-ph/0701231

\bibitem[2001] {L01} Liu, X.-W., Luo, S.-G., Barlow, M.J., Danziger, I.J., Storey, P.J..
2001, MNRAS, 327, 141

\bibitem[2000] {L00} Liu, X.-W., Storey, P.J., Barlow, M.J., Danziger, I.J., Cohen, M.,
Bryce, M., 2000, MNRAS, 312, 585

\bibitem[2006] {M06} Mateus, A., Sodre, L., Cid Fernandes, R.  et al. 2006, MNRAS, 370, 721

\bibitem[2006] {M06} Mathis, H., Charlot, S., Brinchmann, J., 2006, MNRAS, 365, 385
   
\bibitem[1991] {M91} McGaugh, S.S. 1991, ApJ, 380, 140

\bibitem[1979] {P79} Pagel, B.E.J., Edmunds, M.G., Blackwell, D.E., et al. 1979, MNRAS, 189, 95

\bibitem[1967] {P67} Peimbert, M., 1967, ApJ, 150, 825

\bibitem[2004] {P04} Pettini, M., \& Pagel, B.E.J., 2004, MNRAS, 348, L59 (PP04)

\bibitem[2000] {P01} Pilyugin, L.S., 2000, A\&A, 362, 325 

\bibitem[2001] {P01} Pilyugin, L.S., 2001a, A\&A, 369, 594 (P01)

\bibitem[2001] {P01} Pilyugin, L.S., 2001b, A\&A, 373, 56  (P01)

\bibitem[2005] {P09} Pilyugin, L.S. Thuan, T.X., 2005, ApJ, 631, 231 (PT05)

\bibitem[2006] {P06} Pilyugin, L.S. Thuan, T.X., V\'{\i}lchez, J.M., 2006,
MNRAS, 367, 1139

\bibitem[1989] {O89} Osterbrock, D.E., 1989, Astrophysics of Gaseous Nebulae and Active Galactic Nuclei.
Mill Valley, California: University Science Books

\bibitem[2007] {S07} Salim, S. et al. ApJS (in press), arXiv:0704.3611   
  
\bibitem[2006] {S06} Schiavon, R. P. et al., 2006, ApJ, 651, L93

\bibitem[1979] {S79} Seaton, M.J. 1979, MNRAS 187, 73

\bibitem[1983] {S83} Shaver, P.A., McGee, R.X., Newton, L.M., Danks, A.C., Pottasch, S.R.,
1983, MNRAS, 204, 53

\bibitem[1995] {S95} Shaw, R.A., \& Dufour, R.J., 1995, PASP, 107, 896

\bibitem[2006] {Shi06} Shi, F., Kong, X., Cheng, F.Z., 2006, A\&A, 453, 487

\bibitem[1980] {S80} Stasi\'{n}ska, G., 1980, A\&A, 85, 359

\bibitem[2005] {S05} Stasi\'{n}ska, G. 2005, A\&A, 434, 507

\bibitem[2006] {S06} Stasi\'{n}ska, G. 2006, A\&A, 454, L127

\bibitem[2004] {T04} Tremonti, C.A., Heckman, T.M., Kauffmann, G., et al. 2004, ApJ, 613, 898 (T04)

\bibitem[2007] {Y07} Yin, S.Y., Liang, Y.C., Hammer, F. et al. 2007a, A\&A, 462,
535 (Y07)

\bibitem[2007] {Y07} Yin, S.Y., Liang, Y.C., Zhang, B. 2007b, proceedings of "The central
engine of Active Galactic Nuclei" eds. L. C. Ho and J.-M. Wang, astro-ph/0701234

\bibitem[2006] {V06} van Zee, L., Haynes, M.P., 2006, ApJ, 636, 214

\bibitem[1993] {V93} Vila-Costas, M. B. \& Edmunds, M. G. 1993, MNRAS, 265, 119

\bibitem[2007] {W07} Wild, V., Hewett, P. C., Pettini,  M., 2007 , MNRAS, 374, 292

\bibitem[2006] {Z06} Zheng, X. Z., Bell, E. F., Rix, H.-W., Papovich, C.,
Le Floc'h, E., Rieke, G. H., Pérez-Gonzalez, P. G., 2006, ApJ, 640, 784

\end{thebibliography}
\end{document}